\documentclass{aa}

\usepackage{graphicx}

\usepackage{txfonts,textcomp}

\usepackage[colorlinks=true,linkcolor=black,citecolor=blue]{hyperref}

\newcommand{\mvir}{\rm M_{200}}
\newcommand{\rvir}{\rm R_{200}}
\newcommand{\mstar}{\rm M_{*}}
\newcommand{\Msun}{{\rm M_{\sun}}}

\newcommand{\ropt}{\rm R_{opt}}
\newcommand{\cielo}{{\sc CIELO}}
\newcommand{\Mdot}{\rm \dot{M}}
\newcommand{\Vc}{\rm V_c}
\newcommand{\Vmax}{\rm V_{max}}
\newcommand{\hmax}{\rm h_{max}}
\newcommand{\Vr}{\rm V_r}
\newcommand{\mex}{\rm \dot{M}_{\rm ex}}
\newcommand{\mout}{\rm \dot{M}_{\rm out}}

\begin{document} 

   \title{Metal-loaded outflows in sub-Milky Way galaxies in the  ~\cielo~ simulations}

   \author{Valentina P. Miranda
          \inst{1,2}
          \and
          Patricia B. Tissera \inst{1,2}
          \and 
          Emanuel Sillero \inst{1,2}
          \and
          Jenny Gonzalez-Jara \inst{1,2} 
          \and
          Lucas Bignone \inst{3}
          \and
          Ignacio Mu\~noz-Escobar \inst{1,2}
          \and
          Susana Pedrosa \inst{3}
          \and
          Rosa Dom\'inguez-Tenreiro \inst{4}
          }

   \institute{Instituto de Astrofísica, Pontificia Universidad Católica de Chile, Av. Vicuña Mackenna 4860, 7820436, Santiago Chile\\
              \email{valentina.miranda@uc.cl}
         \and
             Centro de Astro-Ingeniería, Pontificia Universidad Católica de Chile, Av. Vicuña Mackenna 4860, 7820436, Santiago Chile.
         \and
             Instituto de Astronomía y Física del Espacio, CONICET-UBA, Casilla de Correos 67, Suc. 28, 1428, Buenos Aires, Argentina.
         \and 
            Departamento de Física Teórica, Universidad Autónoma de Madrid, E-28049 Cantoblanco, Madrid, Spain.}

   \date{Received 1 August 2025 / Accepted 15 December 2025}

\abstract{Supernova (SN) feedback-driven galactic outflows are a key physical process that contributes to the baryon cycle by regulating star formation activity, reducing the amount of metals in low-mass galaxies and enriching the circumgalactic (CGM) and intergalactic media (IGM).}
{We aim to understand the chemical loop of sub-Milky Way (MW) galaxies and their nearby regions.}
{We studied 15 simulated central sub-MW galaxies ($\mstar \leq  10^{10} \ \Msun$) and intermediate-mass galaxies ($\mstar$ $\sim$  10$^{10}$ $\Msun$) from the CIELO-P7 high-resolution simulations. We followed the evolution of the progenitor galaxies, their properties, and the characteristics of the outflows within the redshift range $z = [0, 7]$.  We used two dynamically motivated outflow definitions, unbound outflows, and expelled mass rates to quantify the impact of SN feedback.}
{ At $z \sim 0$, sub-MW galaxies have a larger fraction of their current oxygen mass in the gas phase but have expelled a greater portion beyond the virial radius, compared to their higher-mass counterparts. Galaxies with $M_{*} < \sim 10^{9}~\Msun$ have 10–40 percent of their total oxygen mass within $\rvir$ in the CGM and an equivalent to 10–60 percent expelled into the IGM. In contrast, more massive galaxies have most of their oxygen mass locked by the stellar populations. The CGM of low-mass galaxies predominantly contains oxygen low-temperature gas, which acts as a metal reservoir. We find that the outflows are more oxygen-rich for sub-MW galaxies, $\rm Z_{\rm out}/Z_{ISM} \sim 1.5$, than for higher-mass galaxies, $\rm Z_{\rm out}/Z_{ISM} \leq 0.5$, particularly for $z <2$. Mass-loading factors of $\rm \eta_{out}\sim0-6$ are detected, in agreement with observations. While a weak dependence of $\eta$ on mass and circular velocity is found at $z\sim 0$, a stronger anti-correlation appears for higher redshift.  }
{Our results suggest that sub-MW galaxies may store a significant fraction of metals in their CGM and that the anti-correlation between $\eta$ and stellar mass (or circular velocity) is stronger at $z \sim 2$, which is likely due to a combination of more intense star formation, a higher merger rate, and shallower potential wells.} 


   \maketitle
\section{Introduction}  \label{sec:introduction}
 Galaxy formation and evolution is a complex interplay of physical processes such as the accretion of gas and the interactions and mergers of the central systems with satellites \citep{somerville2015}. 
These processes also regulate the star formation activity, mix the chemical elements in the interstellar medium (ISM), and induce radial migration \citep[e.g.][]{Krumholz2017}. The evolution of stars leads to the synthesis of new chemical elements, which are expelled into the ISM—primarily during the final stages of stellar life—accompanied by significant energy release that can power metal-enriched outflows \citep[e.g.][]{tinsely1980,maiolino2019}.
Supernova (SN) feedback-driven outflows are expected to be one of the major drivers of the baryon cycle \citep{Peroux2020} by reducing the baryon fraction in low-mass galaxies \citep{dave2009} and enriching the circumgalactic (CGM) and intergalactic media (IGM) \citep{christensen2018}. The impact on low-mass galaxies is thought to be stronger due to their shallower potential wells, which facilitate the ejection of material \citep{dekelandsilk1986}. Galactic outflows can also shape the mass-metallicity relation (MZR) of galaxies, as they might be able to transport enriched material out of galaxies, particularly in low-mass galaxies \citep{brooks2007}. Additionally, metal-poor gas inflows are also expected to contribute to the modulation of the global metal content of galaxies and hence their location on the MZR \citep[e.g. ][]{derossi2012,Zenocratti2022,Bassini2024}.
 
 Given the important role that SN feedback plays, studying low-mass galaxies becomes particularly relevant. However, low-mass galaxies are challenging to study because they usually have low surface brightness, which makes them harder to detect \citep[e.g.][]{mcquinn2019}. Therefore, higher-mass galaxies have been more extensively studied. In addition, low-mass galaxies represent a key element in solving cosmological discrepancies in the standard $\Lambda$ Cold Dark Matter Model ($\Lambda$CDM), such as the missing satellites problem or the dark matter distribution in dwarf galaxies \citep{sales2022,kaneshisa2024}. Therefore, low-mass galaxies play a very important role in constraining galaxy formation theories and sub-grid physics modelling \citep{Ma2016,bullock2017}.

 A key parameter to study the impact of outflows in galaxies is the mass-loading factor ($\eta$), which is defined as the rate at which gas is transported out of the system with respect to the star formation rate (SFR). Observations reported $\eta$ ranging from $\sim$ 0.5 to 10 for nearby dwarf galaxies \citep{martin1999,mcquinn2019} and up to $\sim$ 25-60 for extreme starburst systems \citep{heckmann2015,Perrotta2023}. High-redshift estimates of main-sequence galaxies obtained with the Atacama Large Millimeter/Submillimeter Array (ALMA; \citealt{Herrera-Camus2021, Pizzati2023, Birkin2025}) and of low-mass galaxies observed with the James Webb Space Telescope (JWST; \citealt{Carniani2024}) range between $\eta = 0.5$ and 10.
 
 In the Local Universe, observational estimations reported by \citet{mcquinn2019} found only a weak dependency on circular velocity or stellar mass. However, for high redshift galaxies ($z \sim 4-9$) there is evidence of a strong dependency on stellar mass, $\eta \propto M_*^{-0.43}$ and $\eta \propto M_*^{-1.3}$, found by \citet{Pizzati2023} and \citet{Carniani2024}, respectively. These dependencies agree
 with the predictions from the Feedback In Realistic Environments (FIRE) simulations  \citep{Muratov2015,pandya2021}. Additionally, \citet{murray2005} found that the momentum injection by supernovae (i.e. momentum-conserving outflows) into the galaxy scaled as $\eta \propto$ $\Vc^{-1}$, where $\Vc$ is the halo circular velocity, whereas energy-conserving outflows seem to correlate the mass-loading factor as $\eta \propto$ $\Vc^{-2}$ \citep{chevalier1985}. 
 At $z \sim 0$, hydrodynamical simulations often estimate $\eta \sim 10-100$  \citep{vogelsberger2013,ford2014,christensen2018,pandya2021}, which are higher than the observed values, but not so different from the new estimations reported at high redshift. We note that results from numerical simulations depend on the numerical implementation used to model outflows and hence, the analysis and comparison with observations is crucial to set constraints on these models, and to understand the role of and impact played by SN outflows \citep[see also][]{Rosdahl2017, Kim2020, Rocafabrega2021}.
 
 The exploration of the chemical abundances of the ISM and the CGM offers the possibility to further understand the role of SN feedback and the baryon-metal cycle \citep{Peroux2020}. In this context, cosmological simulations of galaxy formation are a powerful tool to investigate the complex interplay between SN feedback and the evolutionary history of galaxies. By tracking the progenitor galaxies and their components, gas and stars, back across cosmic time, simulations enable us to directly study the origin and evolution of outflows, as well as the location of the expelled material.

  Simulations generally find that dwarf galaxies are more efficient than Milky Way (MW)–mass galaxies at ejecting metals due to the lower potential well \citep{scannapieco2008,Christensen2016,Muratov2017,Mina2021}. This is in global agreement with theoretical expectations, which consider the relative importance of the injected SN energy and the binding energy of the system \citet{dekelandsilk1986}. However, there is no clear consensus among different numerical experiments regarding the fraction of metals that remain in the dwarf galaxies and their CGM. For example, the FIRE simulation \citep{Hafen2019} reported that dwarf galaxies could retain between 30 to 100 percent of their metals within all gas inside the virial radius, $\rvir$,  and between 10 and 60 percent in their CGM. In contrast, \citet{christensen2018} reported lower retention rates, estimating 30 to 60 percent for gas within $\rvir$, and 15 to 25 percent in the CGM. Recently, \citet{piacitelli2025} studied a larger simulated sample of dwarf galaxies and found that the warm and cold phases retained between 5-10 percent of metals formed by the dwarf galaxies. They also found a weak correlation between galaxy mass and the CGM metal retention factor. These differences between numerical results might be due to the differences in the subgrid physics and, potentially, to the different methods and modelling assumed to measure outflows as mentioned before. 

In fact, observationally, outflows are often detected using integrated spectra, such as H$_\alpha$ emission lines as tracers of mass loss in galaxies \citep[e.g.][]{martin1999,mcquinn2019,Carniani2024}. In local starburst galaxies, galactic winds are detected using far-UV absorption lines \citep{heckmann2015,chisholm2017}, whereas at higher redshift they are usually detected using rest-frame optical emission lines such as [OII] and [OIII] \citep{Weldon2024} or in the rest-frame far-infrared using the [CII] emission line \citep[e.g.][]{Ginolfi2020,Herrera-Camus2021,Pizzati2023}. Works using simulations generally define outflows as gas particles that have positive radial velocities \citep[see][]{vogelsberger2013,Muratov2017,bassini2022}, or velocities larger than the Bernoulli velocity \citep{pandya2021}. Other works tracked gas particles and measured outflows by following the material that moved beyond the virial radius of a galaxy as a function of redshift \citep{ford2014,angles2017,christensen2018}. Hence, while global trends might agree, more specific differences might be due to the actual definition of outflows.

  In this paper, our main goal is to analyse the impact of SN feedback on the production of outflows and their effect on metal retention in sub-MW galaxies. In particular, we focus on understanding how oxygen is distributed in the different baryonic components in these galaxies and transported via galactic outflows. For this purpose, we studied 15 central galaxies of the high-resolution simulation of the ChemodynamIcal propertiEs of gaLaxies and the cOsmic web (\cielo) project \citep{Tissera2025}. We followed the evolution of the galaxies across their merger trees. We analysed the oxygen abundances of our galaxies, as well as their history of star formation and triggering of outflows.  We explored two dynamically motivated definitions of outflows, one based on the binding energy and the other by tracking particles as they move out of the virial radius \citep[e.g.][]{angles2017}.
  
  This paper is organised as follows. In Section~\ref{sec:simulations}, we provide a brief description of \cielo~simulations,  our galaxy sample, and the main methodology.
  In Section~\ref{sec:outflows} we analyse the selected galaxies and their properties in time, and study the main metallicity relations. In Section~\ref{sec:mass and oxygen distribution}, we analyse for the oxygen distribution in our simulated galaxies, unbound outflows, inflows, and expelled mass. Section~\ref{sec:mass-loading factor} discusses the $\eta$ factor and its dependency on circular velocity and stellar mass at $z\sim 0$. We also study the evolution of $\eta$ in Sect.~\ref{sec:mass-loading factor redshift}. Finally, in Sect.~\ref{sec:discussion and conclusions} we summarise our main results and conclusions. 

\section{The \cielo~simulations and galaxies} \label{sec:simulations}
In this work, we analyse simulated galaxies from the \cielo~project, a suite of zoom-in hydrodynamical cosmological simulations \citep{Tissera2025}.
In Section~\ref{sec:CIELO simulations}, we provide a brief summary of the main characteristics of the \cielo~simulations.  A more detailed description of the initial conditions, subgrid physics, and the main properties of the simulated galaxies can be found in \citet{Tissera2025}. The subset of \cielo~galaxies\footnote{This work utilises the \cielo~galaxy database reported by \citep{Gonzalez-jara2024}.} selected for the analysis is presented in Section~\ref{sec:CIELO galaxies}.

\subsection{The \cielo~simulations} \label{sec:CIELO simulations}

The zoom-in \cielo~simulations were performed by using a version of the {\sc gadget-3} \citep{springel2005}. This version incorporates a multiphase model for the gas component, metal-dependent cooling, a prescription for star formation where stars form in dense, cold gas clumps, and feedback by Type Ia and Type II {supernovae (SNe)  \citep[][]{scannapieco2005,scannapieco2006}. The \cielo~simulations adopt an initial mass function (IMF) by \citet{chabrier2003}, with lower and upper cut-offs of 0.1 $\Msun$ and 40 $\Msun$, respectively.\\
The chemical SN feedback implemented is taken from \citet{mosconi2001}. This model considers 12 chemical isotopes:  \element[ ][1]{H},  \element[ ][4]{He}, \element[ ][12]{C}, \element[ ][16]{O},  \element[ ][24]{Mg},  \element[ ][28]{Si},  \element[ ][56]{Fe},  \element[ ][14]{N},  \element[ ][20]{Ne},  \element[ ][32]{S},  \element[ ][40]{Ca,} and  \element[ ][62]{Zn}. Primordial abundances are assumed for the initial gas particles: ${\rm X}_{\rm H} = 0.76$,  ${\rm Y}_{\rm He} = 0.24$ and $Z =0$, where $Z$ is the metallicity defined as the fraction of elements heavier than He in the baryonic component.

Type II supernova are assumed to form at the final phase of the evolution of stars more massive than  $8 \ \Msun$, and their lifetimes are estimated following \citet{raiteri1996}. \cielo~adopts  the SNII yields by \citet{woosley1995}.
SNIa are assumed to originate from CO white dwarf systems, in which mass transfers from the secondary to the primary start until it outreaches the Chandrasekhar mass. SNIa yields were taken from \citet{iwamoto1999}, and the lifetimes of the progenitors are taken at random within $[0.1,1]$ Gyr \citep{scannapieco2006}. This simple model for the lifetime distributions of the progenitors of SNIa reproduces very well with the results provided by the single-degenerated (SD) model as discussed by \citet{jimenez2015}.

The SN feedback model adopted is able to reproduce galactic mass-loaded winds without introducing mass-dependent scaling parameters. It also includes a multiphase model for the ISM that allows the coexistence of the hot, diffuse phase, and the cold, dense gas phase, where star formation takes place \citep{scannapieco2006,scannapieco2008}. The energy injected by both types of SNe is distributed equally between the cold and hot phases. The energy injected into the cold phase is stored in a reservoir until the gas particles accumulate enough energy to change their thermo-dynamical properties and join the hot phase. The energy injected into the hot phase is instantaneously thermalised. The injection into a cold energy reservoir may delay the effective injection of energy when a galaxy is forming stars at a low rate, but during this period, no outflows are expected. When a gas particle is transformed into stars, the existing energy reservoir is injected into the surrounding gas. The thresholds to separate cold and hot phases have been extensively tested by \citet{scannapieco2005,scannapieco2006}.
 
The initial conditions of \cielo~are consistent with the $\Lambda$CDM cosmological model, with $\Omega_0 = 0.317$, $\Omega_{\Lambda} = 0.6825$, $\Omega_b = 0.049,$ and $h = 0.6711$ \citep{Planck2014}. 
    The initial conditions of the CIELO project are zoom-in regions around target galaxies selected from a dark matter only cosmological simulation. The target haloes were chosen to map different density regions, excluding large groups and clusters. In this work, we use the higher resolution CIELO-P7, centred around a target halo of virial mass\footnote{The virial mass, M$_{200}$, is defined as the mass enclosed within a sphere of radius $\rvir$, at which the mean mass density reaches 200 times the background density.} $\rm M_{200} = 1.3 \times 10^{12}\Msun$.  \citet{Tissera2025} reported that this zoom-in region corresponds to a filamentary structure. The initial mass gas particles have $m_{\rm g} = 3.1 \times 10^{4}\Msun$ and the dark matter particles have $m_{\rm dm} = 2.0 \times 10^{5}\Msun$.

  CIELO-P7 hosts haloes of virial mass within the range, $\mvir \sim 10^{10}-10^{12} \Msun $. The virialised haloes were identified using a friends-of-friends algorithm \citep[FoF,][]{davis1985}, and the substructures within each halo were individualised by using the SUBFIND algorithm \citep{springel2001,dolag2009}. The most massive substructure within a viral halo is classified as a central galaxy. Finally, the merger trees were built using the AMIGA algorithm \citep{knollmann2009}.

 \subsection{The \cielo-P7 galaxies}
 \label{sec:CIELO galaxies}
 We studied 15 central galaxies\footnote{We did not consider three out of the 18 galaxies of \cielo-P7 because they do not show strong SF nor any outflow features.} from \cielo-P7 simulations. We traced the evolution of their progenitor galaxies over 89 snapshots of the simulations,  covering the redshift range $z = [0,7]$. The snapshots have a time separation ($\Delta t$) of a median of $1.6 \times 10^8$ yr with a minimum and maximum of $5 \times 10^7$yr and $1.7 \times 10^8$ yr. This time resolution allows us to effectively measure the impact of the SN feedback on the production of outflows, as we define in Sect.~\ref{sec:outflows}.} We named them according to their SUBFIND IDs at $z = 0$ as shown in Table~\ref{table:fitting_circular velocity}.
 These galaxies have stellar masses within the range $10^8 < \rm M_*/\Msun < 10^{11}$. Hereafter, we denote those with $\mstar<10^{10}$ $\Msun$ as sub-MW galaxies.

 Galaxies were reoriented by aligning the z-axis with the angular momentum vector of the stellar component of each simulated galaxy in our sample. Within this new system of reference, the decomposition of the stellar particles into halo, bulge, and disc was done following the procedure, dubbed AM-E, described by \citet{tissera2012}. This method is based on combining both the binding energy and angular momentum content of stellar particles to define the components. A detailed description can also be found in \citet{Gonzalez-jara2024}.

\begin{figure*}
  \centering
   \includegraphics[width=\textwidth]{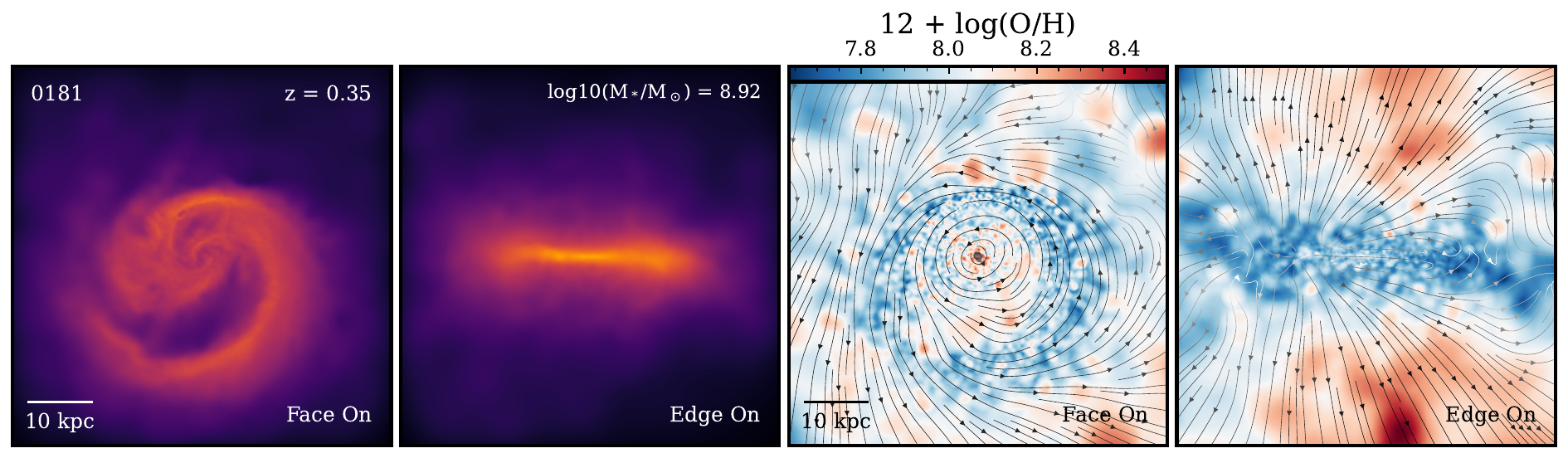}
   \includegraphics[width=\textwidth]{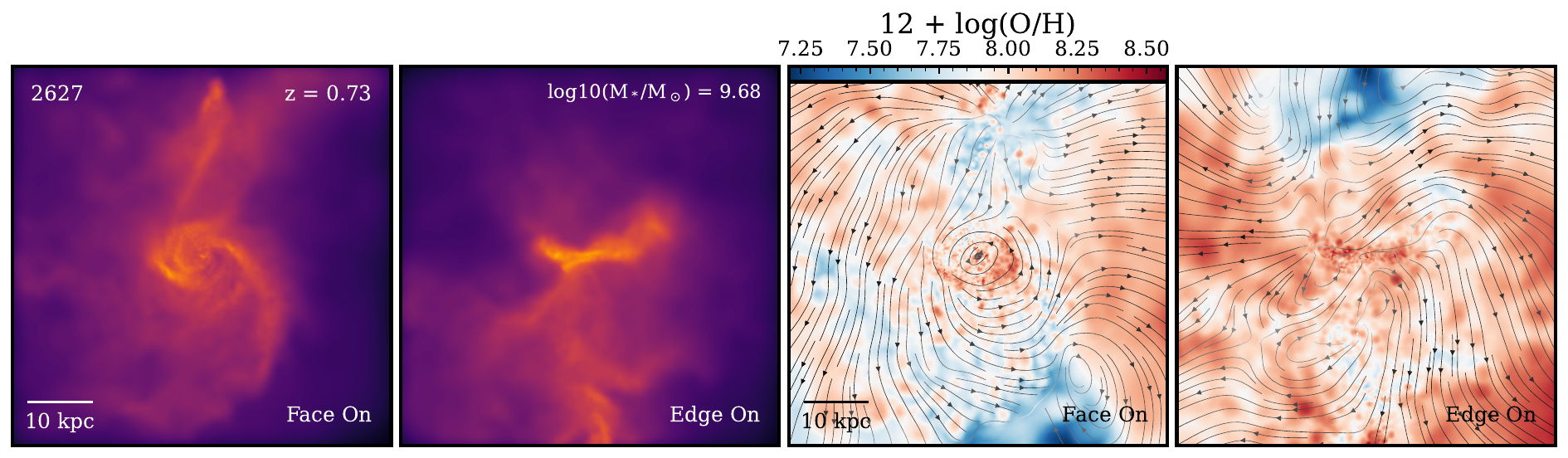}
  \includegraphics[width=\textwidth]{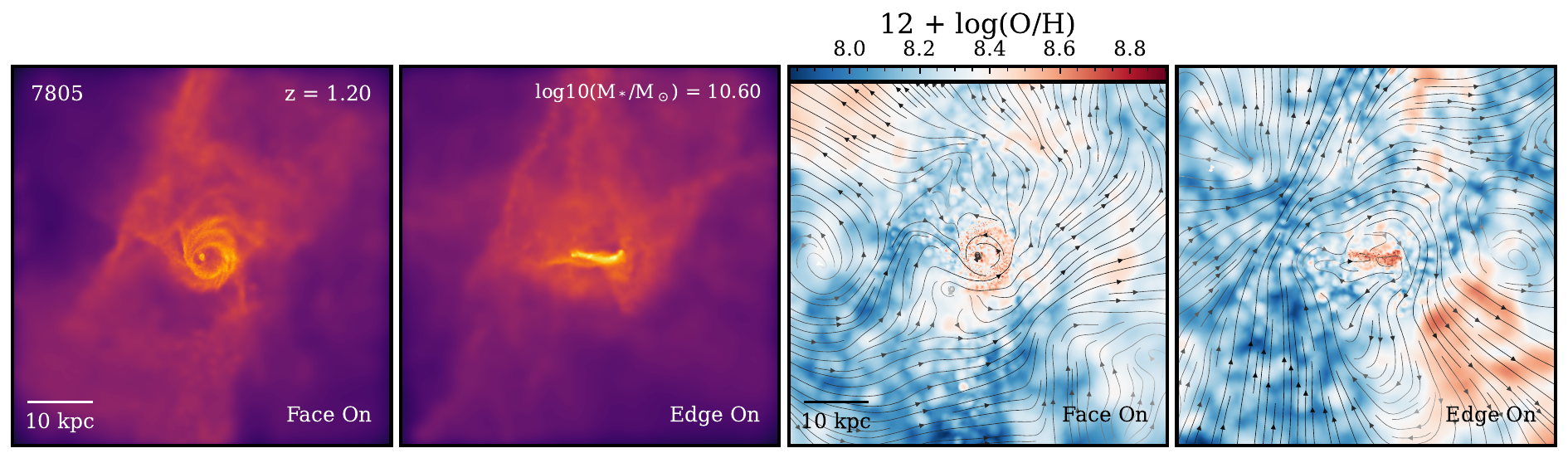}
      \caption{Face-on and edge-on projected gas density (first two panels, respectively) and the corresponding (O/H)  (two right panels) distributions for three selected galaxies of the \cielo~-P7 sample: 0181, 2627, and 7805, taken as examples.
       Each row shows a galaxy at a key stage of its evolution where gas outflows are detected. The $\mstar$ of the galaxies at redshift $z = 0$ is shown in each plot. The streamlines depict the median velocity direction of the gas components. The first two galaxies are more rotationally dominated, while the third galaxy has a complex gas structure resulting from a recent gas-rich interaction. }
         \label{fig:family_picture}
\end{figure*}

 To classify the galaxies morphologically, we define the bulge-to-total stellar mass ratio, B/T, where the total mass is equal to the bulge and disc mass. 
 We clarify that the AM-E method does not allow the detection of bars, and if this component exists, it is included as part of the bulge. Hence, the B/T ratio has to be taken as indicative of the global morphology of the galaxies.
 According to this ratio, most of our galaxies have an important stellar spheroidal component but also show clear rotating gaseous as displayed in Fig.~\ref{fig:family_picture}. 
 
 We confine a galaxy as all gas, stars, and dark matter particles within $1.5\ropt$, where $\ropt$ is defined as the one that encloses 83 percent of the baryonic mass, and within a height set by the maximum height from the disc plane of the stellar particles belonging to the bulge or disc stellar components, $\hmax$. Hence, the ISM is defined as the gas component contained within a galaxy (i.e. $1.5\ropt$ and $\hmax$) while the CGM is defined as the gas that is located within the $\rvir$ and does not belong to the ISM.
 We classify the gas phase and the stellar components along the main branch of the merger trees of each selected \cielo~galaxies from $z=7$ to $z=0$.\\
 \noindent
 As examples in Fig.~\ref{fig:family_picture} we show the face-on (first row) and edge-on (second row) projections for the gas components of the three selected galaxies adopted: 0181, 2627, and 7805. Each galaxy is shown at key stages in the evolution where we detected outflows with the methods described in Sect.~\ref{sec:outflows} (see Fig.~\ref{fig:SFH}). In addition, we show the oxygen abundance maps with streamlines showing the gas velocity direction for the same projections (third and fourth rows, respectively). Galaxies 0181 and 2627 (upper and middle panel) have a well-defined gaseous disc, which exhibits clear signals of gas outflows coming from the central region of the discs and interacting with the CGM (see velocity streamlines). Galaxy 7805, on the other hand, is more massive and has a more dominating stellar spheroidal component. However, as can be seen from this figure, the gas component is also disturbed. 
 We observe that sub-MW galaxies present outflows moving away from the galactic plane in the z-axis. These outflows are oxygen-rich as can be seen in the last two columns of Fig.~\ref{fig:family_picture} (face-on and edge-on projections). 
 From this figure, we can see that outflowing material tends to be more enriched or similarly enriched than the ISM. In contrast, higher-mass galaxies such as galaxy 7805 (lower panel) have a less oxygen-rich outflow than the ISM. To understand the nature of these outflows in Section~\ref{sec:outflows}, we quantify the level of enrichment in relation to the potential well of the systems to analyse the impact of SN feedback.
\\
 
\begin{table*}
\caption{Main physical properties of the selected \cielo~ galaxies $z = 0$.}
\label{table:physicalparameters}   
\centering                  
\begin{tabular}{c c c c c c c c c c c c}       
\hline\hline
Galaxy ID & $\Vmax$ & $\Vc$ & B/T & log(M$_{\rm bar}$/$\Msun$) & log($\mstar$/$\Msun$) & sSFR & 12 + log(O/H)& log($\gamma_{\rm eff}$)\\ 
& km s$^{-1}$ & km s$^{-1}$& & & & $10^{-11}$$\Msun$yr$^{-1}$& dex & \\
\hline   
   2780 & 35.8 & 35.8 & 0.95 & 8.72 & 8.04 & 2.27 & 7.81 & -2.07 \\
   2763 & 39.4 & 35.1 & 0.97 & 8.63 & 8.12 & 8.06 & 8.10 & -2.14 \\
   2736 & 24.3 & 24.3 & 0.96 & 8.41 & 8.17 & 1.42 & 8.42 & -2.15 \\
   9110 & 31.3 & 31.3 & 0.84 & 8.54 & 8.26 & 3.64 & 8.18 & -2.18 \\
   2774 & 42.3 & 42.3 & 0.69 & 8.64 & 8.40 & 7.18 & 8.22 & -2.26 \\
   0200 & 12.8 & 12.8 & 0.79 & 8.73 & 8.68 & 0.22 & 8.36 & -2.67 \\
   0192 & 55.3 & 51.3 & 0.64 & 9.18 & 8.81 & 0.05 & 8.19 & -2.39 \\
   0181 & 70.1 & 70.3 & 0.79 & 9.56 & 8.92 & 0.30 & 8.17 & -2.20 \\
   2717 & 68.7 & 64.7 & 0.45 & 9.48  & 9.11 & 0.68 & 8.34 & -2.38\\     
   2696 & 75.6 & 70.7 & 0.62 & 9.49 & 9.38 & 0.02 & 8.33 & -2.61\\
   2627 & 94.3 & 94.3 & 0.65 & 9.86 &  9.68 & 0.79 & 8.22 & -2.45\\
   0000 & 102.7 & 102.7 & 0.63 & 9.81 & 9.73 & 0.60 & 8.58 & -2.32\\
   8958 & 111.8 & 111.8& 0.63 & 10.21 &  10.17 & 0.02 & 8.50 & -2.98\\
   7805 & 152.7 & 152.7 & 0.79 & 10.60 & 10.60 & 0.32 & 8.81 & -2.18\\
   2389 & 218.1 & 218.1& 0.68 & 10.74 & 10.72 & 0.51 & 8.96 & -2.13\\
\hline                                
\end{tabular}
\tablefoot{Columns from left to right contain the maximum rotational velocity, the so-called circular velocity as defined in Section~\ref{appendix:TFR}, the stellar bulge-to-total ratio, the total baryonic galaxy mass, the total stellar galaxy mass, specific SFR, the oxygen abundance for the star-forming gas, and the effective yields.}
\end{table*}

\section{Outflows, inflows, and star formation rates} \label{sec:outflows}
 In this section, we describe the methods we used to estimate the rate of gas transported in and out of a galaxy and the analysis performed to compare simulations with observations. 
 Regarding the estimation of the outflow rates, there are different approaches in the literature \citep[e.g.][]{vogelsberger2013,Muratov2017,bassini2022}. In this work, we adopt two dynamically motivated definitions. Our first approach defines unbound outflows by taking into account only the gas moving outwards with enough energy to eventually escape the potential well of its host galaxy,
\begin{equation} \label{eq:unbound outflow}
    \Mdot_{\rm out} (t)= \frac{\sum_{\rm i} m_{\rm i,out}}{\Delta t}
,\end{equation}
 where the sum is over the $i$ gas particles, which are unbound and are dominated by outward radial motions such that $\Vr/\sigma > 1$, with $\Vr$ the radial velocity and $\sigma$, the velocity dispersion of the gas particles. We estimated these rates within a $\Delta t$ given by the time interval between two consecutive available snapshots.\\
 The rate $\Mdot_{\rm out}(t)$ was estimated within two radial bins of the CGM: $[1.5 \ropt, 0.5\rvir)$ and $[0.5\rvir, \rvir)$, of a given \cielo~ galaxy. Hence, for each analysed galaxy, $\Mdot_{\rm out}(t)$ was estimated in two radial bins and as a function of time. 
 
 The second approach adopted is based on the determination of the expelled gas particles. These particles are selected as those that reached galactocentric distances larger than the $\rvir$ of a galaxy and never re-entered it at subsequent times, i.e. $r>\rvir$ always remains valid after the particle was expelled the first time. 
 Furthermore, we restrict the gas particles to those with $\Vr/\sigma > 0.5$ from  $z = 7$ to $z = 0$, so that at high redshift,  we only consider the particles with the highest probability of leaving the system. And, for consistency, we maintain this condition down to low values of $z$. 
 The expelled mass rate is defined as 
 \begin{equation} \label{eq:expelled rate}
     \Mdot_{\rm ex}(t) = \frac{\sum_{\rm i} m_{\rm i,ex}}{\Delta t},
 \end{equation}
 where the sum is over the mass of gas particles $i$ that, at a given snapshot, do no longer belong to the system, but at the previous snapshot did (i.e. r$_{\rm i}$ $>$ $\rvir$ with r$_{\rm i}$ being the galactocentric distance of a given particle at a given snapshot). We note that $\Mdot_{\rm ex}(t)$ could also include material associated with galaxy interactions or splash satellite galaxies. Hence, it is an upper-limit estimation of the action of SN feedback.
 
 We also calculated the rate, $\eta$, at which the gas mass is transported out of a system (i.e. $\Mdot_{\rm ex}(t)$ or $\Mdot_{\rm out}(t)$), relative to its SFR,
\begin{equation} \label{eq:mass-loading factor}
    \eta = \frac{\Mdot (t)}
    {\rm SFR}
,\end{equation}
 where the SFR is defined as $\frac{\sum_{\rm i} m_{\rm i,*}}{\Delta t}$ by considering stellar populations with ages younger than $\Delta t = 0.1 \rm Gyr$. We study the mass-loading factors and their relation with physical properties in Section~\ref{sec:mass-loading factor}.

 Additionally, we estimate the rate at which the gas that is not associated to the central galaxy at $z=7$ enters a given galaxy halo at a determined snapshot ($r < \rvir$), but stays within $\rvir$ up to $z=0$. For this purpose, we tracked gas particles in time and calculated the inflow mass rate as
\begin{equation} \label{eq:inflow}
    \Mdot_{\rm in} (t) = \frac{\sum_{\rm i} m_{\rm i,in}}{\Delta t}
,\end{equation}
 where the sum is over the mass of gas particles $i$ that, at a given snapshot, is within the $\rvir$, but in the previous snapshot was not, $r > \rvir$.\\
 Finally, we defined the effective inflow mass rate, $\Mdot_{\rm in}^{\rm eff}$ by applying Eq.~\ref{eq:inflow}, but considering only those infalling gas particles that, at a given time,  reach the ISM of a galaxy (i.e. $r\leq$ 1.5$\ropt$ and z $\leq$ $\hmax$). This rate considers the gas that effectively fell into the ISM of a galaxy and should be considered a lower limit since we do not track the gas particles that fall between snapshots and are transformed into stars.
 
\begin{figure}
      \includegraphics[width=\hsize]{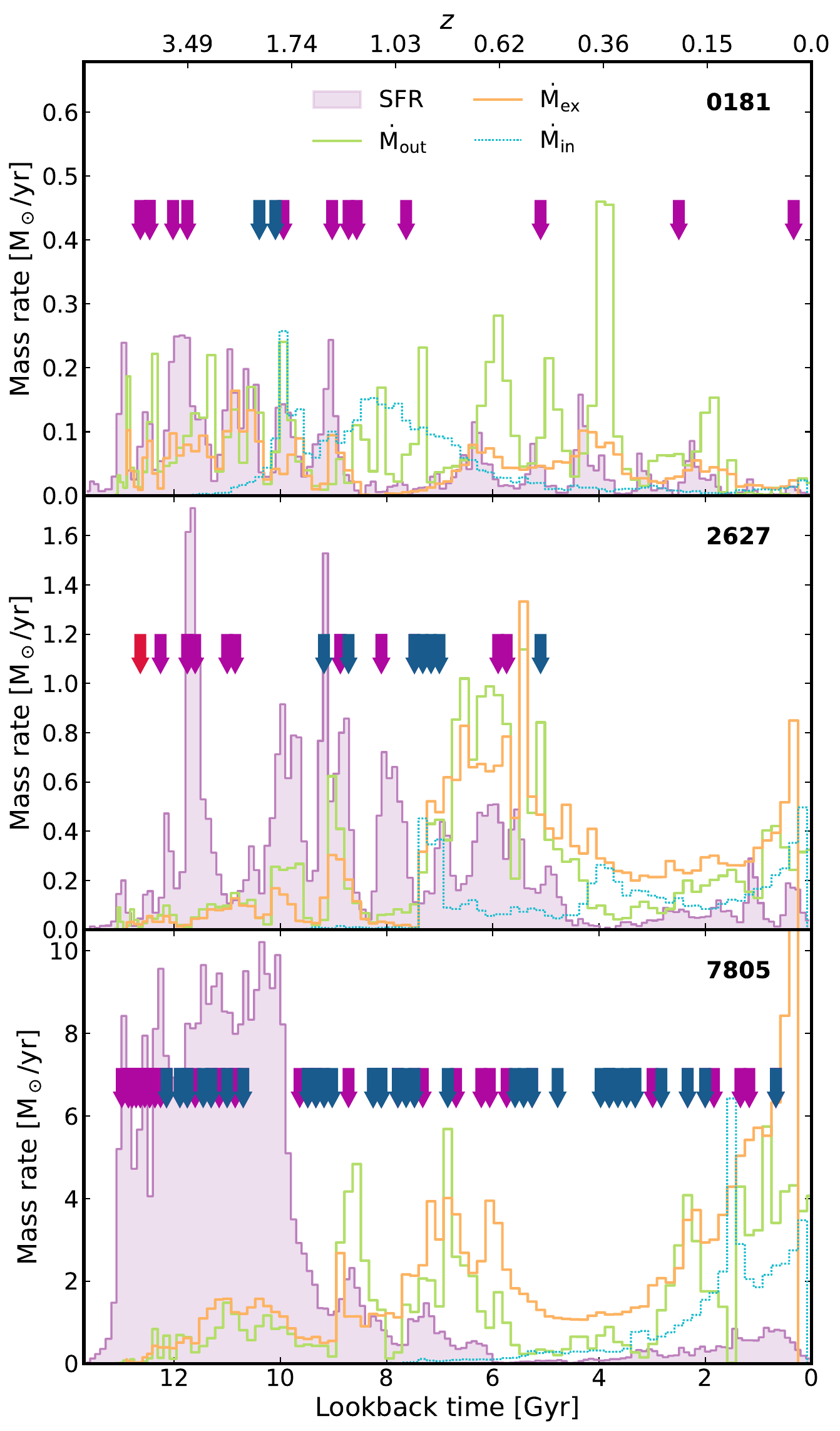}
      \caption{Three galaxies as examples: evolution of SFR (purple shading), $\mout$, the rate of unbound outflow for the inner shell [1.5$\ropt$, 0.5$\rvir$) (solid green lines), $\mex$, the expelled mass rates (solid orange lines), and $\dot{M_{\rm in}}$, the inflow mass rate (dotted cyan line) as a function of lookback time (the inset labels denote the galaxy ID). The infall time of satellites (blue arrows) entering the virial radius and the time of minor and major mergers (pink and red arrows, respectively) are also indicated (see Fig.~\ref{fig:SFH_all} to see the diversity of behaviour in our sample).
              }
         \label{fig:SFH}
\end{figure}

 As an example, in Fig.~\ref{fig:SFH}, we display the SFR, the unbound outflows for the first shell $[1.5\ropt, 0.5\rvir)$, expelled mass rate, and infall mass rates evolution for a sample of three of our galaxies\footnote{We show galaxies 0181, 2627, and 7805, just for simplicity as they are good representatives of our sample. A complete picture is shown in Fig.~\ref{fig:SFH_all}.}
 As can be seen from Table~\ref{table:physicalparameters}, our galaxies have very low star formation activity and some of them are quenched at $z \sim  0$. Only galaxies 2627 and 7805 had recent small starbursts. However, their specific SFR, defined as  ${\rm sSFR = \frac{\rm SFR}{\mstar}}$, are in good agreement with observations of nearby galaxies of similar mass  \citep{cedres2021}.
 In general, all analysed galaxies show bursty star formation histories, albeit with differences in strength and number.
 We can appreciate this in  Fig.~\ref{fig:SFH}, where we show examples of common behaviours. Galaxy 0181 (as well as 2780, 2763, 2736, 9110, 2774,  0200, 2696, 2717, and 2389, shown in Fig.~\ref{fig:SFH_all}) has several starbursts throughout its evolution. Galaxy 2627 (and 8958, shown in Fig.~\ref{fig:SFH_all}) has multiple starbursts of comparable strength across time. 
 Galaxy 7805 (as well as 0192 and 0000, shown in Fig.~\ref{fig:SFH_all}) undergoes a strong and more extended starburst followed by a series of later weaker bursts.

 As can be seen from Fig.~\ref{fig:SFH}, starbursts are typically followed by a decrease in star formation as the energy released by the SN heats the surrounding gas in conditions of forming stars. The injection of energy can trigger gas outflows of different strengths. In Fig.~\ref{fig:SFH} we also display $\mout$ and $\mex$. As can be seen after the bursts there is often an increase of these rates, indicating the triggering of an unbound outflow and an increment in the expelled mass.
 In fact, from Fig.~\ref{fig:SFH} it can be appreciated that the peaks of the unbound outflow rates are slightly delayed with respect to the peaks of the SFR comparing with the expelled mass rates\footnote{We only show the first shell for simplicity as both shells behave similarly.}. This is due, in part, to the time required for the outflow, produced within the ISM, to reach the CGM. We recall that  $\mout$ is measured in concentric shells as defined in Eq.~\ref{eq:unbound outflow}. The peaks of the expelled mass rate agree better in some cases with the starbursts. This can be related to the action of interacting or back-splash galaxies. It has been extensively shown that there is an increase of star formation activity with decreasing pair-wise galaxy separation \citep[e.g.][]{lambas2003}. This behaviour is reproduced by simulations \citep[e.g.][]{katzandhernquist1999, tissera2001,torrey2012,bignone2019,rodriguez2022}. These starbursts then induce the SN feedback, driving galactic outflows. In Fig.~\ref{fig:SFH} we also depicted the times of galaxy infall defined when the satellite enters the virial radius of the central galaxies (blue arrows) and the times of mergers, defined as the time when the SUBFIND algorithm no longer recognises the satellites as separate systems for minor mergers (pink arrows) and major mergers (red arrows). We define major mergers as those in which the stellar mass ratio of the satellite galaxy to the central galaxy exceeds $\mu_* = 0.25$ galaxies, and minor mergers as those with $\mu_* < 0.25$. For this analysis we consider satellite galaxies with a baryonic mass larger than $10^6\Msun$ at the time of infall.
 
  In agreement with previous works, we detect that after an interaction or merger with a satellite galaxy, a burst in star formation is often produced, presumably fuelled by the accretion of gas from the satellite. SN events take place soon after the star formation episode, releasing energy into the ISM, which can trigger outflows.  However in some cases, the satellite strips gas from the central galaxy during interaction that could be detected as expelled material. An example of this is galaxy 7805, which exhibits an expelled mass rate around 5 Gyr ago, with no SFR counterpart. This is caused by the fact that the expelled mass rate method might include gas particles associated to the tidal stripping of the satellites at larger galactocentric distances. While this caveat adds noise to our estimations, it does not significantly affect our conclusions. In fact, this situation might also be present in observations when galaxies had a close interaction.\\

It should be noted that each galaxy has a unique formation history shaped by its interactions, mergers, an inflow, and outflow of gas. Therefore, although we summarise the main trends, every system exhibit a distinct evolutionary path, resulting in a diversity of behaviours. This can be seen in more detail in Fig.~\ref{fig:SFH_all} for the remaining \cielo-P7 galaxies. Overall, we found that their SFRs are modulated by gas-rich mergers/interactions, as well as the SN feedback-triggered outflows in agreement with previous results.

\subsection{The metallicity of gas outflows and inflows}\label{sec:metallicity}

\begin{figure}
   \centering
   \includegraphics[width=\hsize]{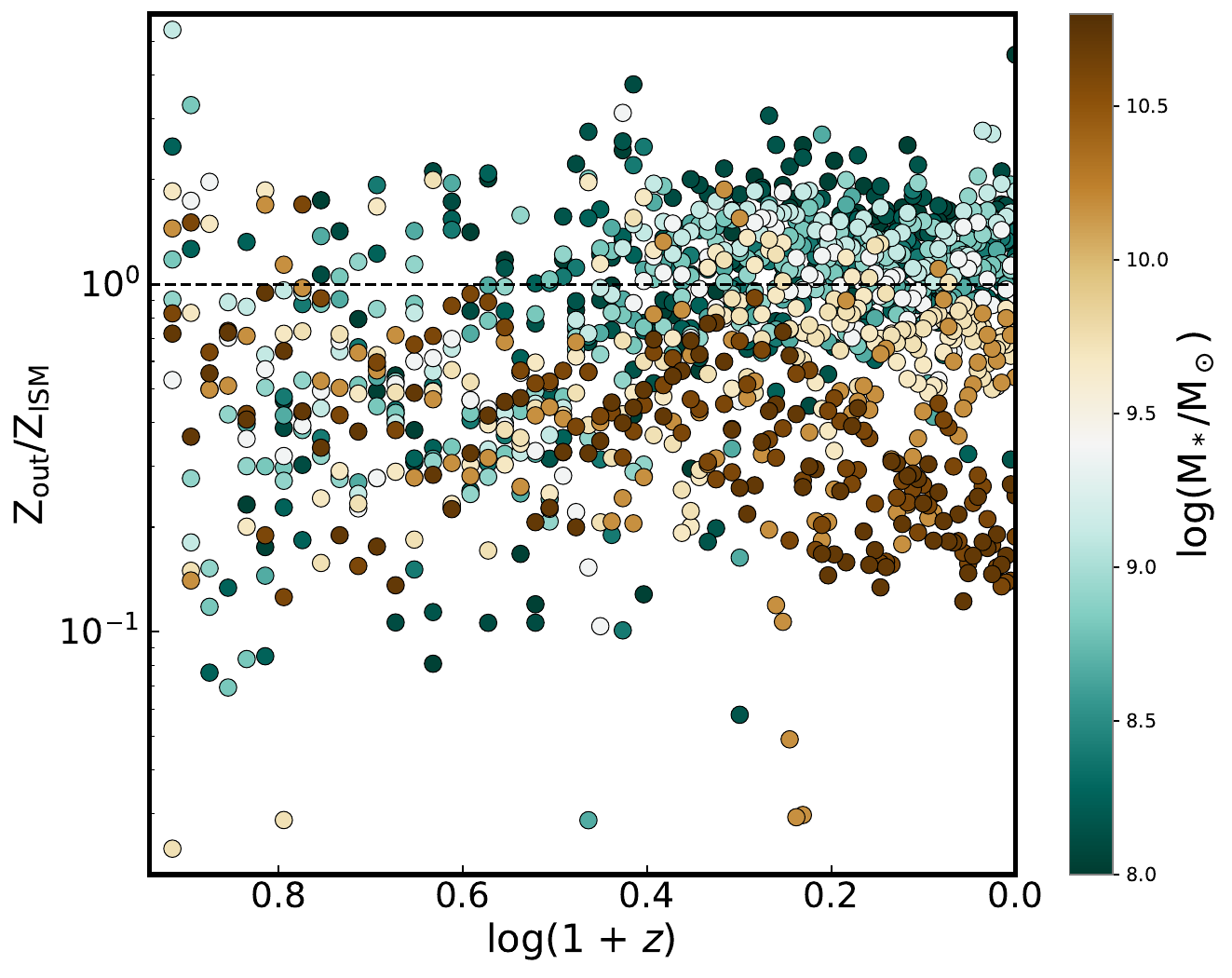}
      \caption{Ratio of the unbound outflow metallicity to the ISM metallicity, $Z_{\rm out}/Z_{\rm ISM}$, for the analysed galaxy and their progenitors as a function of redshift. The dashed black line shows $Z_{\rm out}/Z_{\rm ISM} = 1$.
      The colour-code denotes  $\mstar$ at $z = 0$.
      }
         \label{fig:Zout}
\end{figure}

 During starbursts, gas outflows are expected to transport metals from the galaxy into the CGM \citep{Peroux2020}. Observational studies support this, finding that the expelled material is enriched with respect to the global ISM \citep{cameron2021, hamel2024,mishra2024}. 
 In order to study the level of enrichment of the simulation outflows, we estimated the ratio between the metallicity for the unbound outflows within the $[1.5\ropt, \rvir)$ shell and the ISM metallicity, Z$_{\rm out}$/Z$_{\rm ISM}$ at each available snapshot\footnote{We note that there are more snapshots covering $z < 1$ as explained in Section 2.}. We performed this calculation for each galaxy and its progenitors. To visualise the evolution in time of this ratio and its dependence on stellar mass, in the upper panel of  Fig.~\ref{fig:Zout}, we show the galaxies coloured by $\mstar$ of the central galaxies at $z=0$. We note that the progenitors have smaller stellar masses, and hence, smaller potential wells, as one moves to higher redshift.  We use the stellar mass as a reference to assess the impact of SN feedback to produce chemically enriched outflows with respect to their own ISM or gas inflows.
 
 As can be seen Fig.~\ref{fig:Zout},  the outflow metallicity becomes similar to or larger than the ISM metallicity from redshift $z \sim$ 2. By this redshift, our simulated galaxies have already had several star formation episodes.
 At $z >2 $ regardless of the $z=0$ stellar mass, their lower mass progenitor have outflows which tend to be less enriched than their ISM, except for some systems.  We calculated an average of Z$_{\rm out}$/Z$_{\rm ISM} = 0.56 \pm 0.34$ for the progenitors of massive galaxies, and Z$_{\rm out}$/Z$_{\rm ISM} = 0.65 \pm 0.61$ for the progenitors of sub-MW galaxies for $z >2 $. For $z\leq 2$ the progenitors of massive galaxies tend to retain more efficiently their metals, with an average of Z$_{\rm out}$/Z$_{\rm ISM} = 0.39 \pm 0.25$, whereas the progenitors of the sub-MW galaxies have an average of  Z$_{\rm out}$/Z$_{\rm ISM} = 1.13\pm 0.47$. Outflows are clearly more efficient to transport metals out in sub-MW systems and their progenitors during the whole analysed redshift range. 
 The weaker trend detected at higher redshift could reflect the fact that all progenitors are relatively small systems, similarly susceptible to SN feedback. At lower redshift, the progenitors of massive galaxies become increasingly more massive and therefore less affected by SN feedback, which helps to make the trend more pronounced.

 These results are consistent with previous studies from \citet{christensen2018} where for simulated dwarf and spiral galaxies, an anti-correlation with the stellar mass and a peak in the $Z_{\rm out}/Z_{\rm ISM}$ ratio around $z \sim 2$  were reported. Also, \citet{Muratov2017} found that outflows in lower-mass galaxies ($\mstar = 10^7-5\times10^{9} \Msun$) are in general more metal-rich than their ISM where $Z_{\rm out}/Z_{\rm ISM}$ $\sim 1-1.5$, similar to what we obtain in this work.

Since metals expelled from a galaxy can later be re-accreted, and gas inflows can penetrate the virial radius \citep[e.g.][]{ceverino2017, collacchioni2019}, we also estimated the ratio between the inflow metallicity and the ISM metallicity.
We found that $Z_{\rm in}$ is lower than $Z_{\rm ISM}$ for all analysed galaxies as a function of time. For $z > 2$, we estimated for high-mass systems $Z_{\rm in}/Z_{\rm ISM}= 0.16 \pm 0.19,$ whereas for the sub-MW galaxies, an average of $Z_{\rm in}/Z_{\rm ISM} = 0.02 \pm 0.06$. 
For $z \leq 2$, we found that  higher-mass systems have $Z_{\rm in}/Z_{\rm ISM} = 0.15 \pm 0.17$, whereas the sub-MW galaxies,  $Z_{\rm in}/Z_{\rm ISM}= 0.18 \pm 0.32$.
There is a large scatter in our data, suggesting that the accretion of gas also includes pristine and recycled material from the CGM and IGM. Unlike Z$_{\rm out}$/Z$_{\rm ISM}$, we find no clear dependence on stellar mass or redshift.
 
 \subsection{Mass-metallicity relation} \label{sec:MZR}
 
\begin{figure}
   \centering
   \includegraphics[width=\hsize]{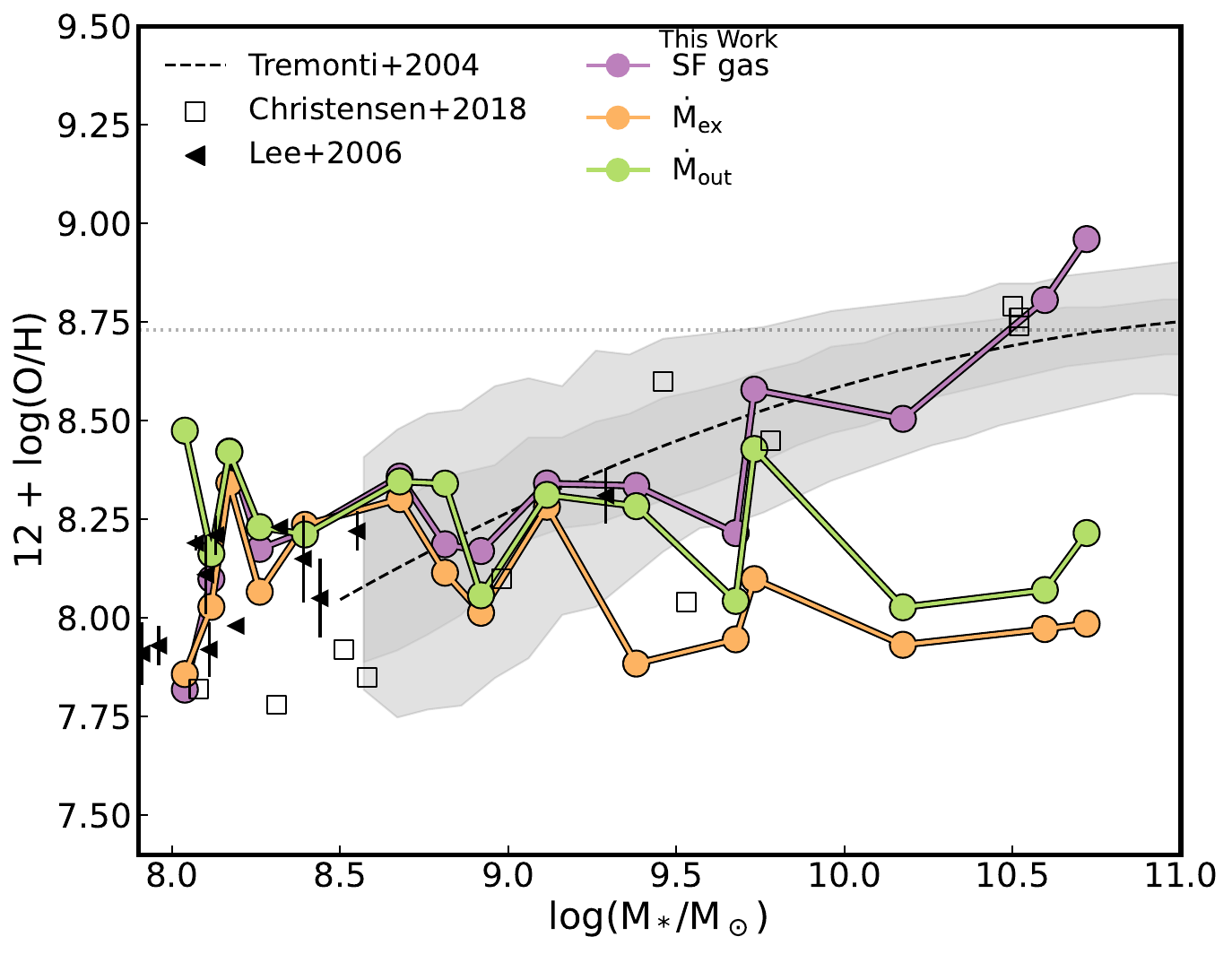}
      \caption{Mass-metallicity relation for unbound outflows in the first radial interval of the CGM (green circles), for the expelled mass rates (orange circles), and for the star-forming gas (purple circles) for each analysed galaxy. Solid lines are only included to facilitate the visualisation. Observations from \citet[][leftwards-pointing triangles]{lee2006} and simulations from \citet[][black squares]{christensen2018} are shown. The MZR relation by \citet{tremonti2004} scaled to solar values of \citet[][dotted line]{lodders2019} is shown in dashed black line. The 16-84 percentiles (dark grey contours) and the 2.5-97.5 percentiles (light grey contours)  scaled from \citet{tremonti2004} are also displayed.}
         \label{fig:mzr}
\end{figure}

 We have already shown that SN feedback can trigger metal-loaded galactic outflows, able to transport material outside of the \cielo~galaxies into the CGM and even the IGM. This redistribution of chemical elements can also affect the global metallicity of galaxies and hence, the MZR \citep{brooks2007}. 
 Therefore, to study the impact of outflows and inflows on the metal content of our galaxies at $z = 0$, we also estimated the MZR \citep[see also][]{Tissera2025}. In Fig.~\ref{fig:mzr}, we display the MZR for the star-forming gas (i.e. gas with instantaneous $\rm SFR > 0$ within the simulated galaxies; purple circles). For comparison, we also calculated the oxygen abundance for the unbound outflows in the first radial bin as previously defined in Section~\ref{sec:outflows} (green circles) and for the expelled mass rates (orange circles). We also include observational and other simulated results\footnote{Observations and model by \citet{tremonti2004} have been rescaled to the adopted solar values  $12 + \rm log(O/H)_\odot = 8.73$ \citep{lodders2019} at $\mstar =$ $10^{11} \Msun$.}.

 As can be seen from Fig.~\ref{fig:mzr}, the SF regions in the analysed galaxies reproduce a MZR in agreement with observations \citep[see also][]{Tissera2025}. The simulated values are within the observed range reported by \citet{lee2006}  with a median scatter of 0.15 dex for the sub-MW galaxies and of 0.12 dex for the higher-mass galaxies with respect to \citet{tremonti2004} relation.
The expelled particles and the unbound outflows determine a weaker relation with stellar mass than the metallicity of the SF gas.
 For high-mass galaxies, the unbound outflows are less oxygen-rich than the SF regions, while the expelled material has even lower enrichment. However, the level of enrichment of the SF regions and the expelled/outflow material is more similar for sub-MW galaxies\footnote{ It should be noted that Fig.~\ref{fig:mzr} shows the SF gas within the ISM, which only accounts for a part of the total ISM component. }. 

\subsection{Effective yields}

\begin{figure}
   \centering
   \includegraphics[width=\hsize]{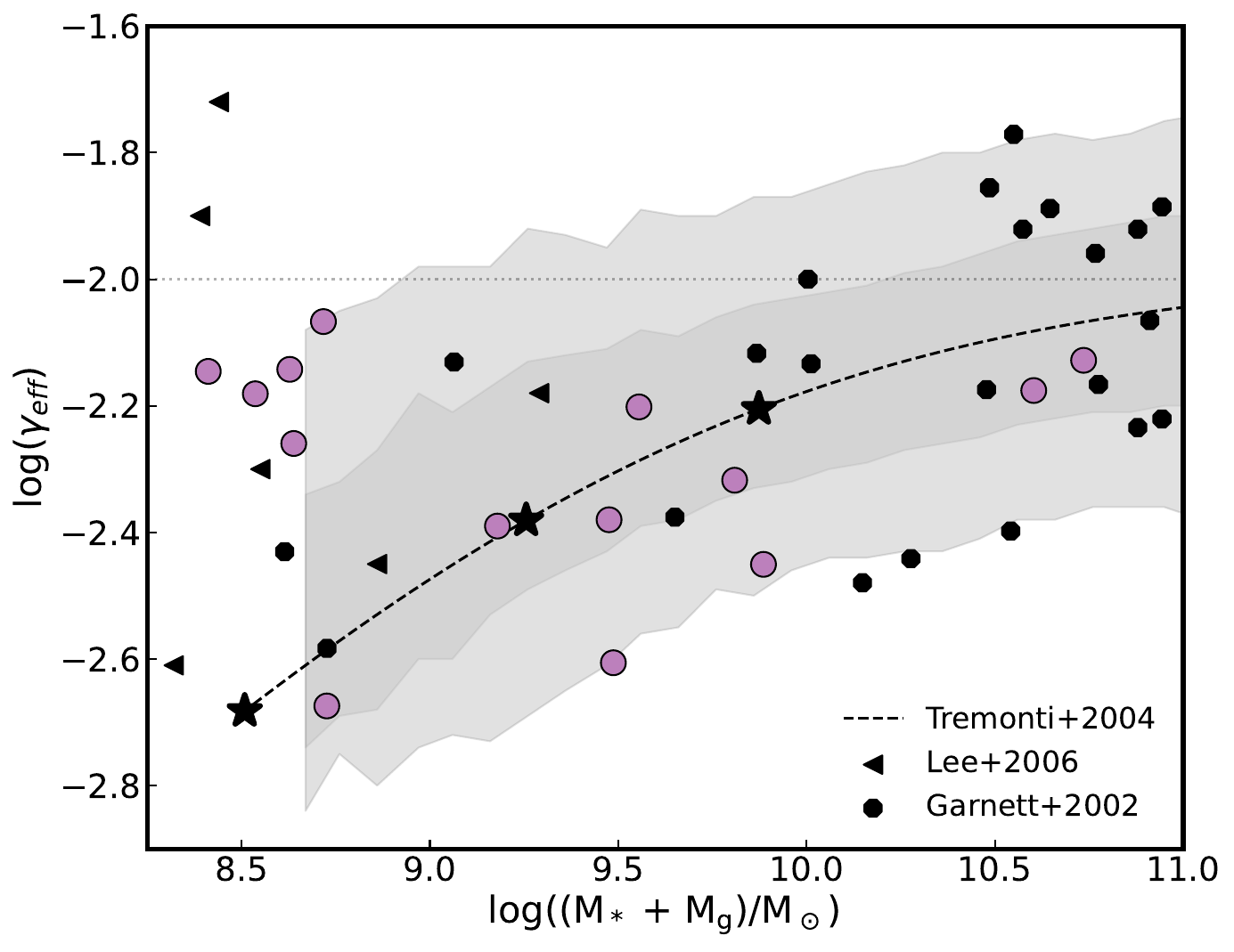}
      \caption{Effective yields as a function of baryonic mass for our galaxies (purple circles). For comparison, observational results reported by \citet[][leftwards-pointing black triangles]{lee2006} and \citet[][black octagons]{Garnett2002} are also included. 
      The shaded areas denote the 16-84 percentiles (dark grey) and the light grey contour shows the 2.5-97.5 percentiles from observations of \citet{tremonti2004}. Additionally, we show the analytical model fitted by these authors \citet[][dashed black line]{tremonti2004}. Black stars show the location of systems that have experienced different levels of metal loss: 80, 60, and 40 percent.
              }
         \label{fig:yields}
\end{figure}

 According to the closed box model \citet{talbot1971}, where no inflows or outflows are taken into account, and instantaneous recycling and mixing are assumed, the true stellar yield can be expressed as a function of the gas fraction $\mu$ and metallicity Z. Effective yields are obtained based on measured s Z and $\mu$, such that
 
\begin{equation}
    \gamma_{\rm eff} = \frac{Z}{\ln{(1/\mu)}}
,\end{equation}
 where $Z$ is defined in Sect.~\ref{sec:simulations} and $\mu$ is given by the gas mass in the ISM divided by the total baryonic mass $\mu \approx \frac{\rm M_{\rm g}}{\mstar + M_{\rm g}}$. 
In a pure closed-box model, a galaxy would tend to have a true yield, $\gamma$, independently of the initial gas mass. As a reference, it is usually included the solar value, $\gamma_\odot = 0.01$\footnote{The solar value was obtained from \citet{tremonti2004}.}. 
The effective yields allow us to estimate how much a galaxy departs from the close box approximation \citep{dalcanton2007}. In Fig.~\ref{fig:yields} we show the effective yields estimated for the simulated \cielo~galaxies (purple circles) and for observations of nearby dwarf irregular galaxies from \citet{lee2006} and spiral and irregular galaxies from \citet{Garnett2002}.

 From Fig.~\ref{fig:yields} we appreciate that none of our galaxies behaves as perfect closed boxes as expected. However, there is a correlation between  $\gamma_{\rm eff}$ and the baryonic mass, where for higher-mass galaxies, the effective yields are larger. This is consistent with sub-MW galaxies experiencing more metal removal from their ISM. The estimated $\gamma_{\rm eff}$ values are given in Table~\ref{table:physicalparameters}.
 We also incorporated the estimates reported by \citet{tremonti2004} by employing an analytical model that includes metal ejection by SN outflows. This model was fitted to the observed data, and the star symbols indicate systems with varying degrees of metal loss. Most of our simulated galaxies are consistent with this model, which predicts progressively larger metal losses in lower-mass systems. However, we identify five sub–MW systems in our simulations that do not exhibit the expected level of metal loss but are within the observed range. This can be attributed to their lack of strong starbursts and, consequently, weaker outflows. We note that there are observational data with similar behaviour \citep{lee2006}. Hence, larger observational and simulated data are needed to better constrain galaxies at the low-mass end, which seem to show a large diversity of behaviours in both simulations and observations \citep{sales2022}.

 As discussed in Section \ref{sec:metallicity}, inflows are dominated by low-metallicity gas that can dilute the ISM. 
Hence, we also estimated  MZR of the gas inflows to assess the level of enrichment of the accreted gas, finding that the inflows have lower metallicity as shown in Fig.~\ref{fig:MZR_in} at $z =0$. A median of 12 + log O/H $=6.95^{7.28}_{6.16}$ for the sub-MW galaxies and of 12 + log O/H $= 7.64^{7.68}_{7.43}$ for the higher-mass galaxies (upper and lower number represent the $75^{\rm th}$ and $25^{\rm th}$ percentiles).
 Outflows tend to be metal-enriched and remove chemical elements from a galaxy. Both processes contribute to lowering the effective yields. As a result, our measurements of the effective yields show some scatter around the observed relation. While the ISM of low-mass galaxies is less-enriched following the MZR, these systems have a significant fraction of gas, producing similar effective yield to massive systems. Overall, we note that there is no clear trend between the effective yields and the total baryonic mass.\\
 
\section{Mass and oxygen distribution at z = 0}\label{sec:mass and oxygen distribution}
\begin{figure}
   \centering
   \includegraphics[width=\hsize]{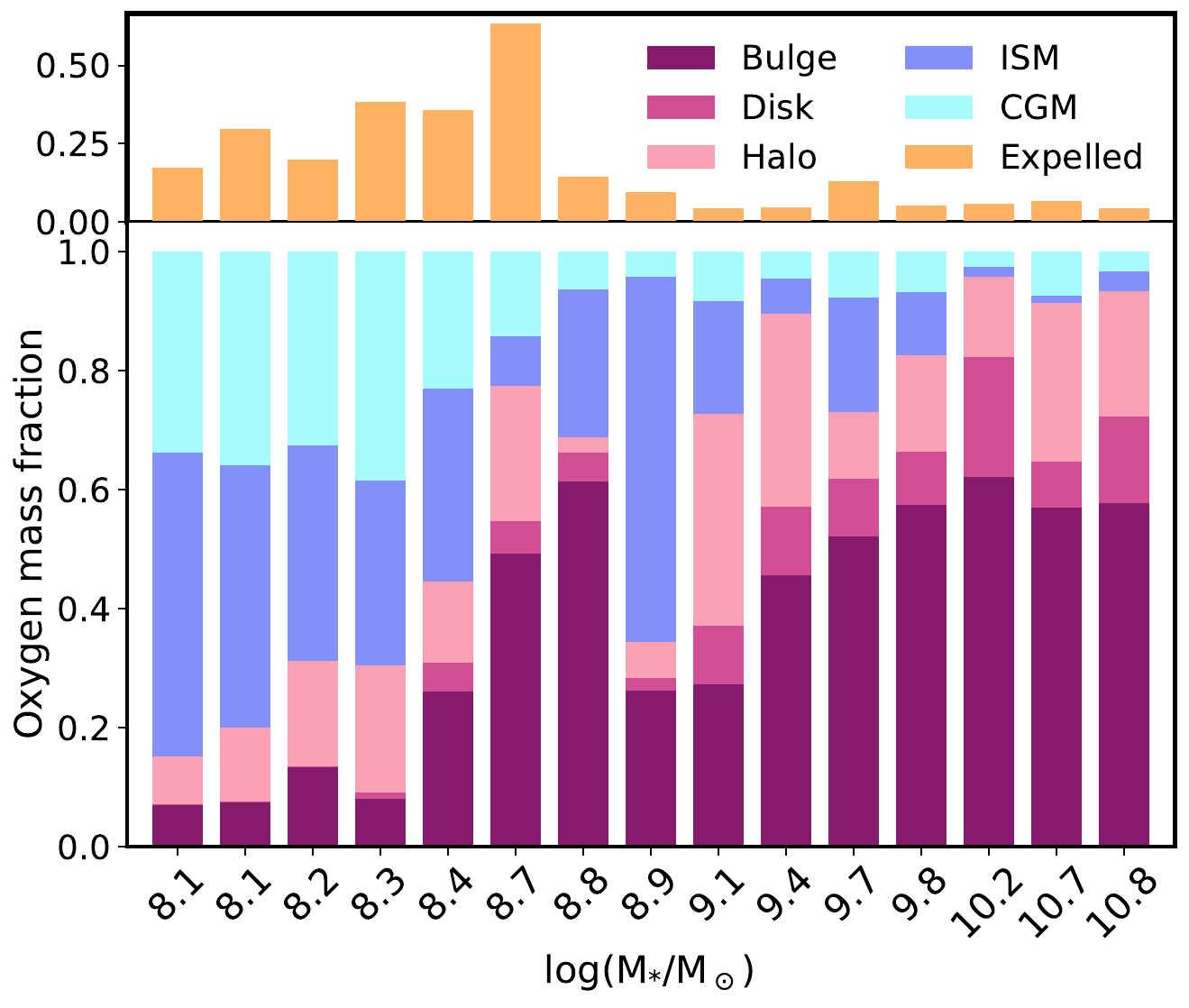}
     \caption{Lower panel: Oxygen mass fraction for stellar and gas components for our simulated galaxies. Upper panel: Expelled oxygen gas mass fraction to the total oxygen mass in a given galaxy. Galaxies are organised by increasing stellar mass from left to right.}
         \label{fig:oxygen_fraction}
\end{figure}

 In this section, we analyse the distribution of oxygen in the different stellar and gaseous components of our galaxies to assess the metal cycle triggered by SN feedback. In the main panel of Fig.~\ref{fig:oxygen_fraction} we summarise the oxygen mass distribution per baryonic component for our galaxies. The oxygen mass fractions for each galaxy have been normalised to its total oxygen mass at $z=0$. Hence, this figure shows the metal budget distribution to the different stellar and gaseous components. Galaxies are ordered according to increasing stellar mass from left to right. For each analysed galaxy, we show the fraction of the oxygen mass in the  ISM (purple), the CGM (cyan), and the stellar mass components of the bulge (burgundy), the disc (dark pink), and the stellar halo (light pink). Additionally, the upper panel displays the expelled fraction (orange) defined as the total expelled mass, i.e. the sum of the mass of all expelled gas particles in a range of redshift $z = [0,7]$, normalised to the total oxygen mass within the virial radius at $z = 0$. It provides an estimation of the oxygen mass lost with respect to the current oxygen mass present within $\rvir$.

 We find that, in high-mass galaxies,  the stellar components store about 90 percent of the total oxygen mass. In particular, the bulge of massive galaxies contains a significant amount of oxygen, which decreases rapidly with decreasing stellar mass to account for about 30 percent in the less massive galaxies, $\sim 10^8 \Msun$.  Consistently, in these systems the oxygen mass fraction in the ISM and CGM increases. However,  there are variations among galaxies of similar masses,  which reflect the different evolutionary and star formation histories. We find that, for galaxies with $M_{*} < 10^{8.7} \rm M_{\odot}$, the CGM gathers between 10-40 percent of the oxygen found with $\rvir$, while for higher mass galaxies, this fraction is less than 10 percent.

  As can be seen in the upper panel of Fig.~\ref{fig:oxygen_fraction}, the expelled oxygen mass increases for the decreasing stellar mass galaxies, implying that sub-MW galaxies lose more oxygen than higher-mass systems. Higher-mass galaxies are more efficient at forming stars, locking metals into them. SN driven-outflows are not expected to affect them strongly due to their deeper potential wells and hence, facilitating the retention of metals, as we discuss in the next section. Conversely, lower-mass galaxies have shallower potential wells; therefore, it is easier for the material to escape the galaxy. We estimate that for sub-MW galaxies, an equivalent to $\sim$ $10-60$ percent of the current oxygen mass was expelled into the IGM, whereas for higher-mass galaxies, this fraction represents less than 10 percent. These estimations agree with previous studies \citep[][see also Sect.~\ref{sec:introduction}]{Muratov2017,Hafen2019}.  

\begin{figure}
   \centering
   \includegraphics[width=\hsize]{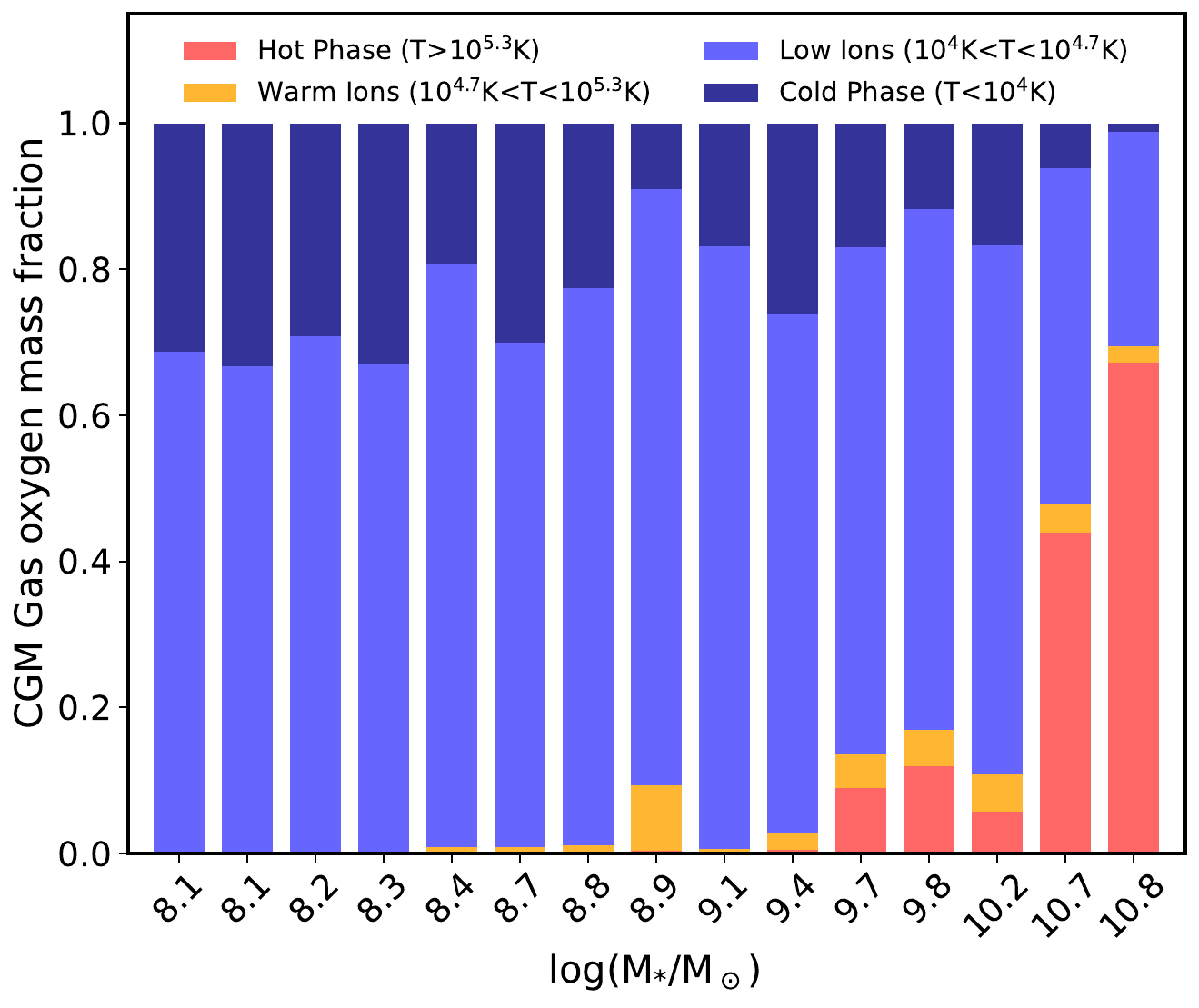}
    \caption{Oxygen mass fractions in the CGM displayed by gas phase defined according to temperature ranges given by \citet{peeples2014} and \citet{Muratov2017} (see inset labels). Galaxies are sorted by increasing stellar mass from left to right.}
    \label{fig:oxygen_temp}
\end{figure}

 In Fig.~\ref{fig:oxygen_temp} we show the breakdown of the oxygen mass stored in the different phases of the CGM. We define different CGM phases according to the gas temperature as hot phase (T $>10^{5.3}$K), warm ions ($\rm 10^{4.7} K < T <10^{5.3} K$), low ions ($\rm 10^{4}K < T <10^{4.7}K$) or cold phase ($\rm T<10^{4} K$), following previous works \citep{peeples2014,Muratov2017,Tumlinson2017, rocafabrega2019}. 
 Sub-MW galaxies have a larger fraction of oxygen in low-ion gas phase. Hot phase gas stores a larger fraction of oxygen with increasing stellar mass, as it reaches about $\sim 70$ percent in the CGM of the most massive galaxy in our sample (see Table~\ref{table:oxygen fraction}). On the other hand, the cold phase gas contributes about $\sim 30$ percent of the CGM in sub-MW galaxies, agreeing with \citet{Muratov2017}. This is not unexpected since as we move to lower-mass haloes, the virial temperature, $\rm T_{\rm vir}$, also decreases, and the temperature threshold to define the low ions and cold phases are closer to the virial temperature of these small haloes (see Table~\ref{table:galaxy properties}).
 Nevertheless, there are variations in the fraction of metals in the hot phase for galaxies with similar masses, which can be attributed to their different evolutionary paths and star formation histories.

\section{Mass-loading factors}\label{sec:mass-loading factor}
 In this section, we analyse the mass-loading factors, $\eta$, and compare them with observations and previous numerical works at $z=0$, $z=1$ and $z=2$. 
The low number statistics both in simulations and observations prevent us from performing detailed statistical analysis. Instead, we estimated the median values of the  \cielo~ simulated $\eta$ in the two defined mass intervals: sub-MW and massive galaxies. In the following analysis median $\eta$ and corresponding 25-75th percentiles are also included.

 \subsection{Local mass-loading factors} \label{sec:circular velocity}
 
 Following \citet{Muratov2017}, we computed a redshift-averaged value over $z = [0, 0.5]$ to enable a better comparison with observations\footnote{We note that our sub-MW galaxies, except for galaxy 2627, do not have a recent strong star formation episode.}. This leads to an artificial high $\eta$ at $z \sim 0$ (see Eq.~\ref{eq:mass-loading factor}) as has also been pointed out in previous works \citep{pandya2021}. 
 As a result, estimating $\eta$ from just one snapshot is not feasible. Hence, we obtain the $\eta$  as defined in Eq.~\ref{eq:mass-loading factor} by integrating the SFR and corresponding outflows for $z \leq 0.5$ (5 Gyrs). We applied the same definition of Section~\ref{appendix:TFR} at all analysed redshifts to estimate $\Vc$.
 
\begin{figure}
   \centering
   \includegraphics[width=\hsize]{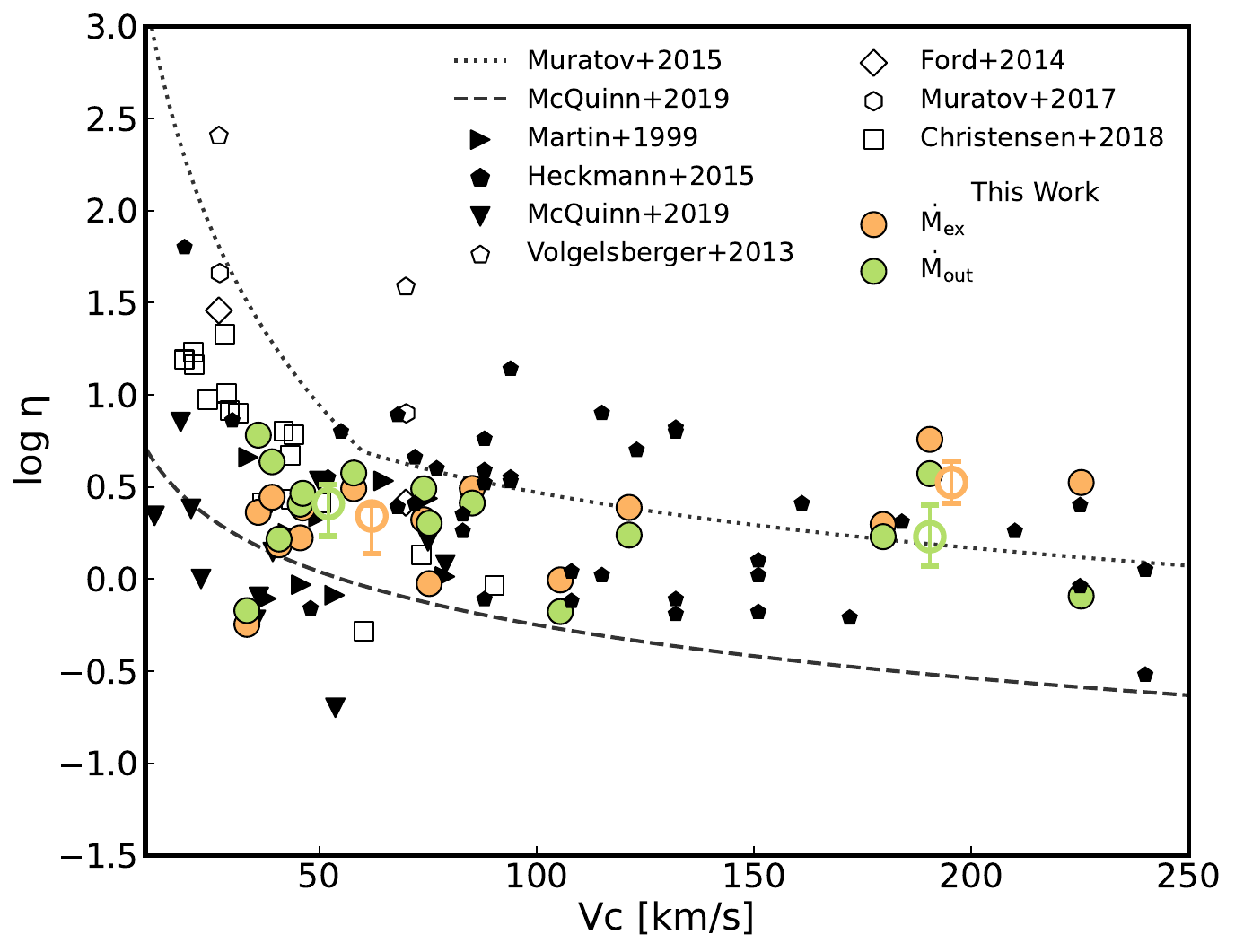}
      \caption{Mass-loading factors as a function of $\Vc$ for \cielo~galaxies for $z$ $\leq$ 0.5, using the unbound outflow (green circles), and the expelled mass rates (orange circles).  Observational data of dwarf, spiral, and starburst galaxies from \citet[][rightwards-pointing black filled triangles,]{martin1999} and of nearby dwarf galaxies from \citet[][downwards-pointing black solid triangles,]{mcquinn2019} are shown for comparison. Additionally, numerical estimations from \citet[][black open pentagons]{vogelsberger2013}, \citet[][black open diamonds]{ford2014}, \citet[][black filled hexagons,]{Muratov2017} and \citet[][black open squares]{christensen2018}. The best fitting regression for the $\eta \propto$ $(1+z)^{1.3}\Vc^{-\alpha}$ with $\alpha = 3.2$ for $\Vc \leq 60 \ {\rm km s^{-1}}$ and $\alpha = 1$ for $\Vc > 60 \ {\rm km\rm s^{-1}}$  at $z = 0.25$ reported by \citet[black, dotted line][]{Muratov2015} and for the $\eta \propto$ $\Vc^{-0.96}$ relation reported by \citet[][dashed black lines,]{mcquinn2019} is also depicted. Median values for  $\eta_{\rm out}$ (green open circles) and for $\eta_{\rm ex}$ (orange open circles) in two mass bins, sub-MW and high-mass galaxies are shown with the corresponding 25-75th percentiles. For visualisation purposes the median values of $\Vc$ for the expelled mass rate measurements have been artificially moved $+10$km/s.}
         \label{fig:eta_vc_z05}
\end{figure}

 In Fig.~\ref{fig:eta_vc_z05} we display $\eta$ as a function of $\Vc$, for the unbound outflows ($\eta_{\rm out}$) measured in the first radial bin of CGM and for the expelled mass rate ($\eta_{\rm ex}$). 
 Estimations for a variety of simulations and observations are also displayed for comparison.

 Our mass-loading factor estimations range between $\eta_{\rm out}~\sim 0.7-6 $ for the unbound outflows as shown in Fig.~\ref{fig:eta_vc_z05} (green circles). The specific values are displayed in Table~\ref{table:mass-loadingfactor}.
 This is consistent with observational estimations reported for dwarf galaxies by \citet{mcquinn2019}. To quantify a potential trend, we estimated the median values of $\eta_{\rm out}$ and $\eta_{\rm ex}$. We find that sub-MW galaxies have a median of  $\log_{10}(\eta_{\rm ex}) =  0.34^{0.40}_{0.14}$ dex for the expelled mass rate and of  $\log_{10}(\eta_{\rm out}) = 0.41^{0.51}_{0.23}$ dex for the unbound outflows, whereas for high-mass galaxies the median value for the mass-loading factors for the expelled mass rate is $\log_{10}(\eta_{\rm ex}) = 0.52^{0.64}_{0.41}$ dex and of $\log_{10}(\eta_{\rm out}) =  0.23^{0.40}_{0.07}$ dex for the unbound outflows.

 The \cielo~galaxies follow the general observational trend, whereby higher circular velocities correspond to lower mass-loading factors, although the correlation is weaker.
 Indeed, a Spearman correlation analysis shows that the simulated relations do not seem to depend on $\Vc$ as can be seen from Table~\ref{table:fitting_circular velocity}. Nevertheless, in this table we provide the best-fitting parameters that yield $\alpha \sim 0.3$, consistent with observational results by \citet{mcquinn2019}. It should be noted that observations for our mass range also appear to have a weak correlation, whereas for lower masses the dependence on $\Vc$ seems to be stronger. As seen in Fig.~\ref{fig:eta_vc_z05}, for the expelled mass rates (orange circles) we obtain slightly higher values, $\eta_{\rm ex}$ $\sim$ 0.6-5.7  \citep[e.g.][]{forster2019,mcquinn2019,concas2022}, but
  consistent with estimated mass-loading factors reported by other numerical works \citep[e.g.][]{ford2014,christensen2018}.

\begin{figure}
   \centering
   \includegraphics[width=\hsize]{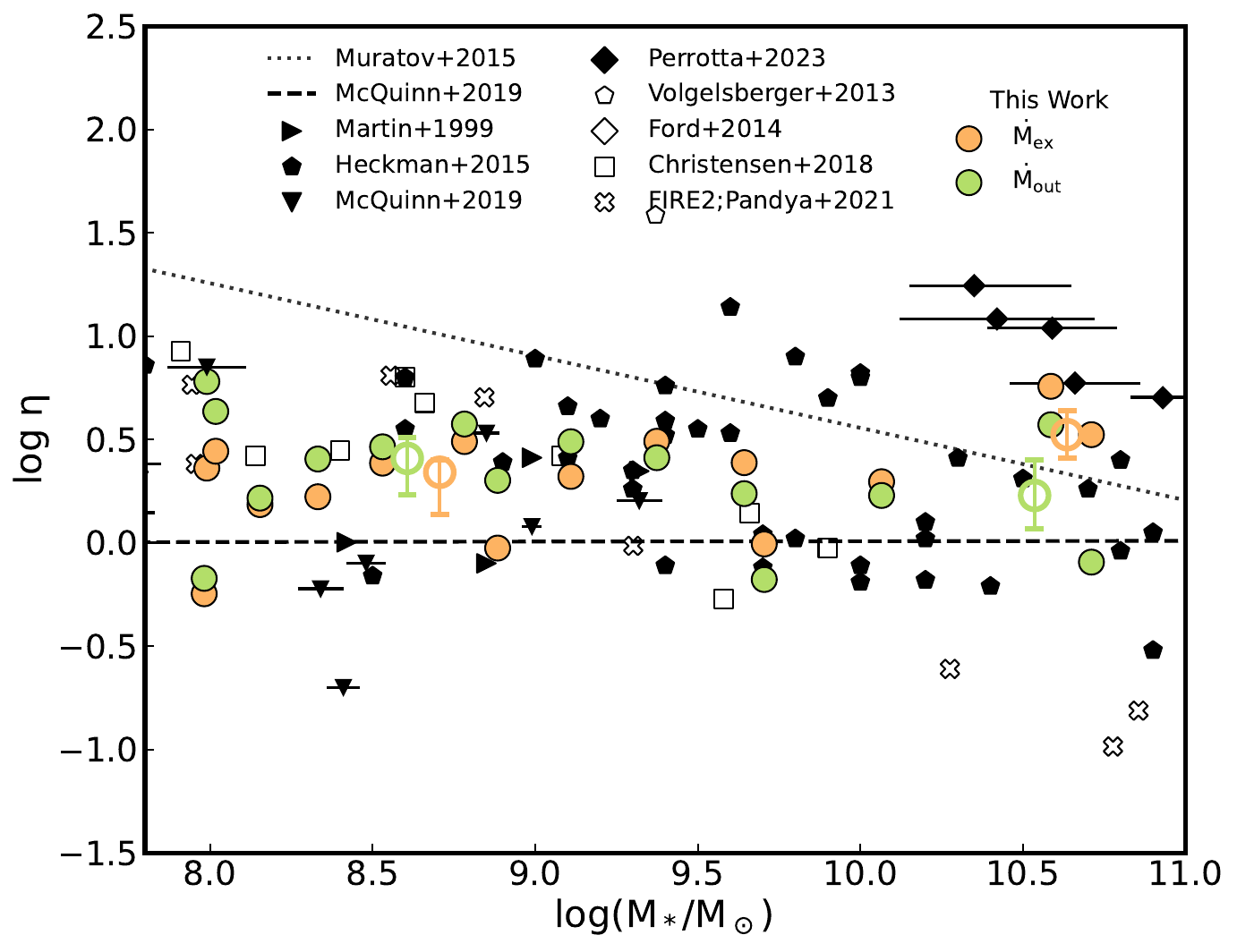}
      \caption{Mass-loading factors as a function of stellar mass using the unbound outflows and expelled mass rates. The reported relation by \citet{Muratov2015} $\eta \propto \mstar^{-0.35}$. Symbols are described in Fig.~\ref{fig:eta_vc_z05} except for FIRE-2 simulation measurements by \citet[][black cross,]{pandya2021} and for observational estimations for massive, compact, and starburst galaxies at $z = 0.4 - 0.7$ by \citet[][black diamonds]{Perrotta2023}. For visualisation purposes the median values of $\mstar$ have been artificially moved $-0.05$ dex for the unbound outflows measurements, and $+0.05$ dex for the expelled mass rate.}
         \label{fig:eta_mstar_z05}
\end{figure}

 In Fig.~\ref{fig:eta_mstar_z05} we present the mass-loading factors as a function of stellar mass. We show the same observations and simulations for comparison previously mentioned in Sect.~\ref{sec:circular velocity}, along with the FIRE-2 simulations by \citet{pandya2021} and observational data of massive, compact, and starburst galaxies at $z = 0.4$–$0.7$ from \citet{Perrotta2023}. The best fitting relations,  $\eta \propto$ $\mstar^{0.04}$ from \citet{mcquinn2019}, and $\eta \propto$ $\mstar^{-0.35}$ from  \citet[][FIRE-1]{Muratov2015} are also shown.

 As can be seen from  Fig.~\ref{fig:eta_mstar_z05}, the simulated $\eta$ obtained for the  \cielo~galaxies agree with observations. We note that some numerical simulations reported a stronger correlation with stellar mass \citep{vogelsberger2013,ford2014}, and some of them show mass-loading factors between one and two orders of magnitude higher than observations for low velocity systems, except for FIRE-2 measurements by \citet{pandya2021}. \citet{mcquinn2019} found a very weak negative correlation with stellar mass. As can be seen from Table \ref{table:fitting_circular velocity}, we find no significant correlation.

 \subsection{Mass-loading factor for z > 0} \label{sec:mass-loading factor redshift}
To study how outflows impact galaxies at different redshift, we study the evolution of the mass-loading factor with redshift. We consider two redshift intervals of $z = 0.5-1.5$ $(z \sim 1)$ and $z = 1.5-2.5$ $(z \sim 2)$. These redshift intervals provide us information on a particular interesting period of galaxy formation, from the cosmic noon when star formation is more active and hence, the impact of SN feedback is more significant.  We followed the same procedure applied for the estimation of the $\eta$ at $z \sim 0$. 

\begin{figure*}
   \includegraphics[width=0.5\textwidth]{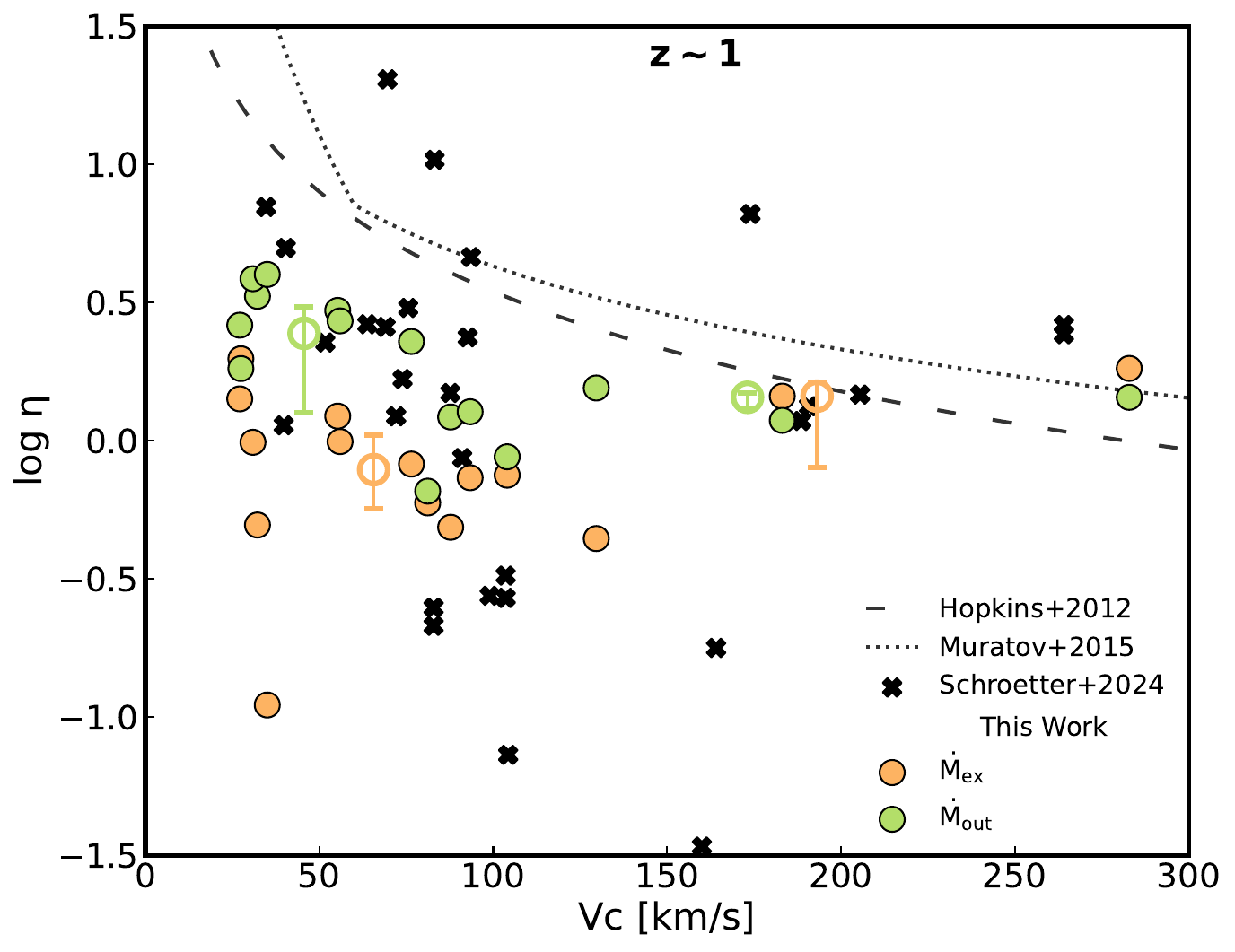}
    \includegraphics[width=0.5\textwidth]{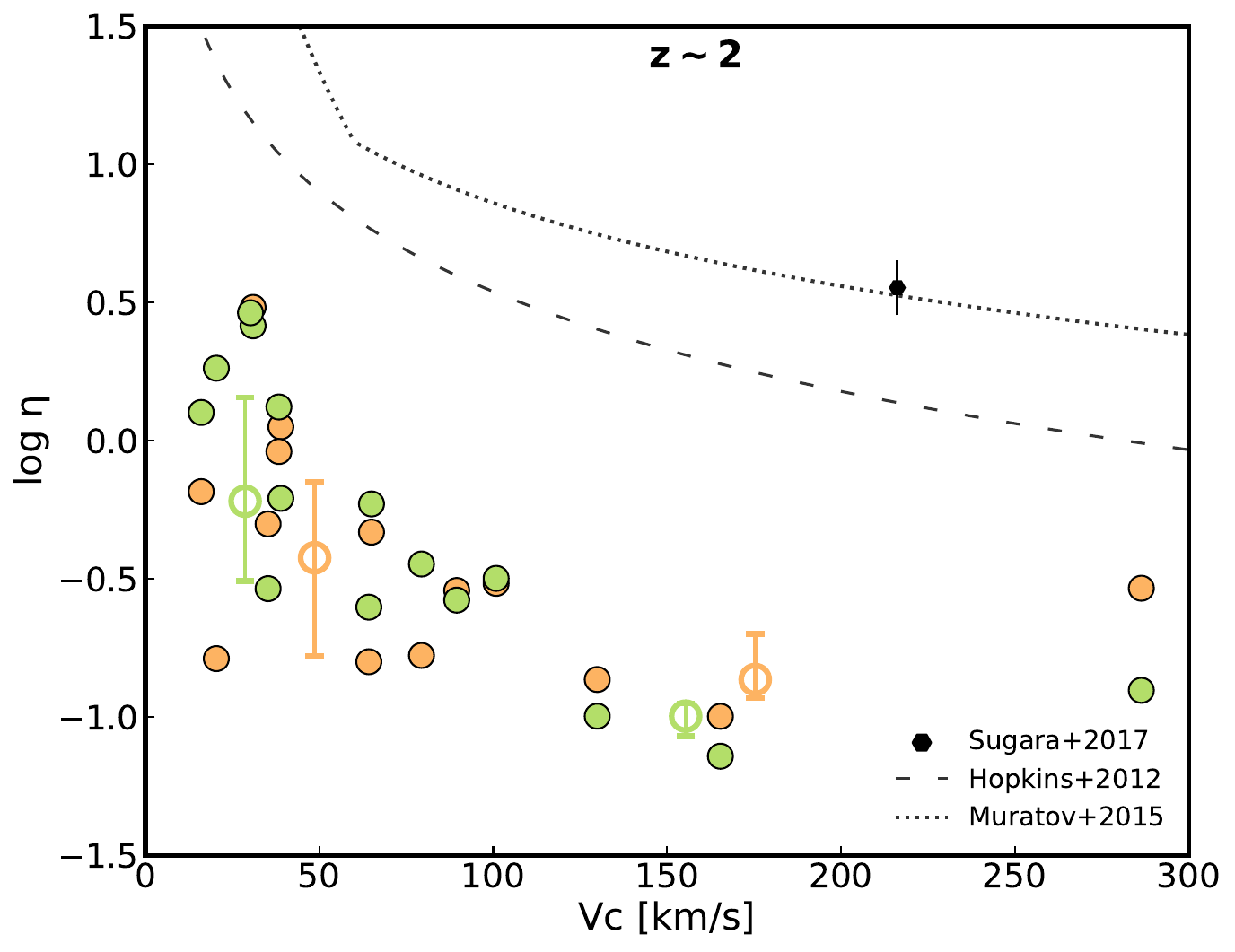}
     \includegraphics[width=0.5\textwidth]{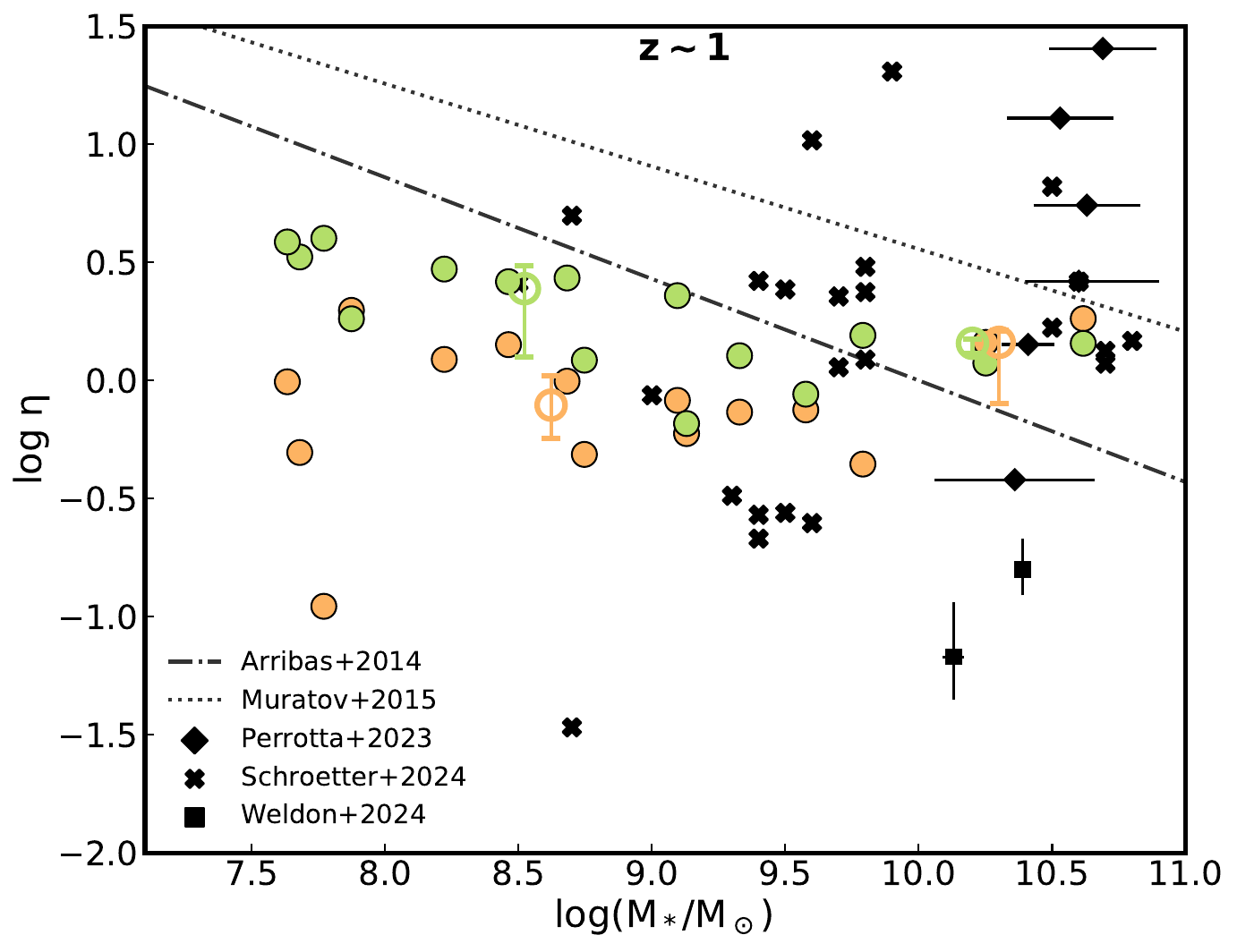}
      \includegraphics[width=0.5\textwidth]{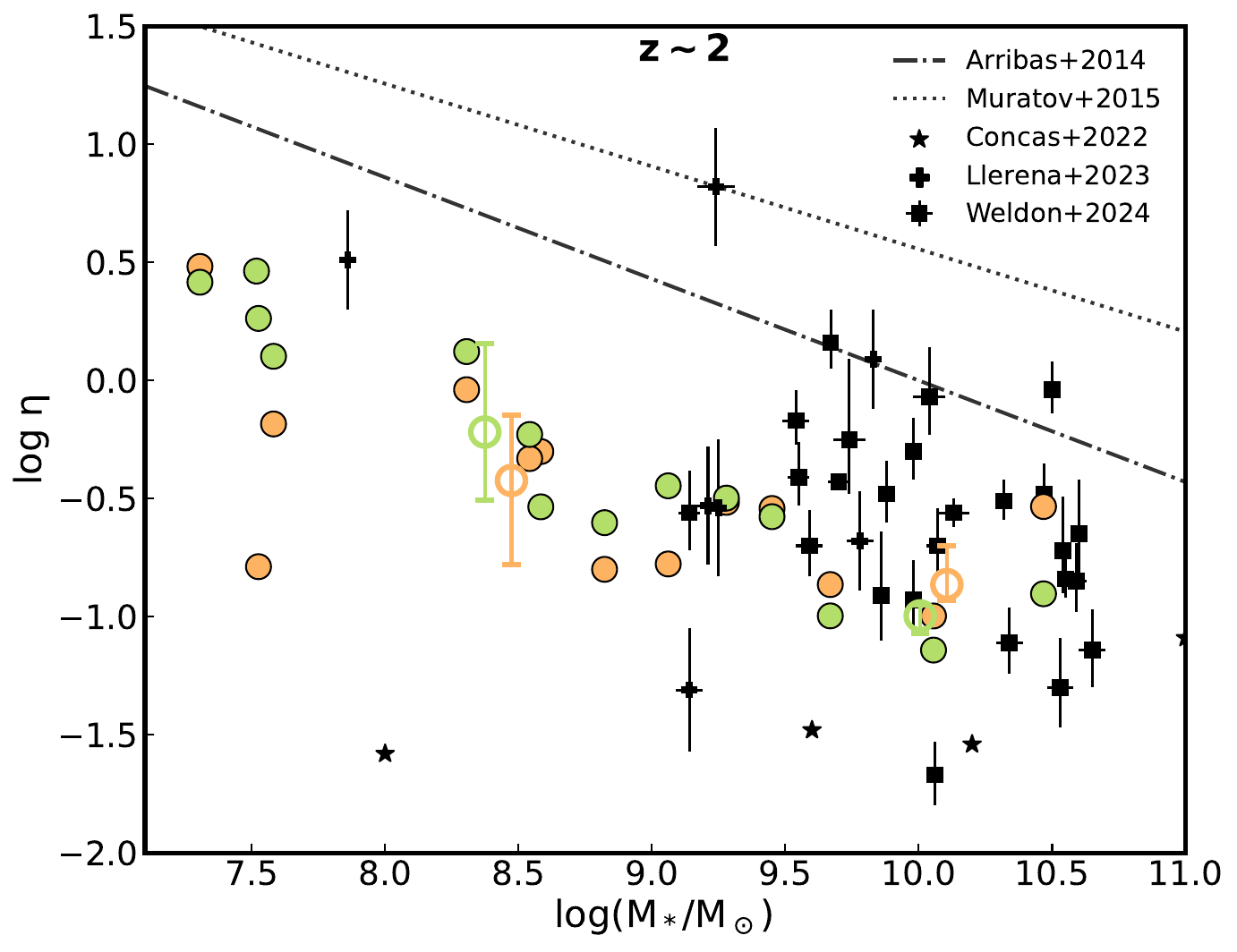}
      \caption{Mass-loading factor, $\eta$, as a function of circular velocity, $\Vc$ (upper-panel) and stellar mass, $\mstar$ (bottom-panel). Symbols and solid lines for this work and \citet{Perrotta2023} are previously described in Fig.~\ref{fig:eta_vc_z05}. Observational data of star-forming galaxies from MEGAFLOW survey \citep{Schroetter2016} by \citet[][black cross,]{Schroetter2024} and of star-forming galaxies from the MOSFIRE survey by \citet[][black squares,]{Weldon2024} are shown, both at $z\sim1$. Observations at $z\sim2$ of a star-forming galaxy from \citet[][black hexagon]{Sugahara2017}, of main-sequence galaxies from the KLEVER survey by \citet[][black stars,]{concas2022} and low-mass star-forming galaxies by \citet[][black cross]{Llerena2023} are also displayed. The best-fitting regression reported by \citet{Muratov2015} previously described in Fig.~\ref{fig:eta_mstar_z05}, $\eta \propto \Vc^{-1.2}$ for the relation reported by \citet[][black dashed line]{hopkins2012} for $z \sim 1$ and  $\eta \propto \mstar^{-0.43}$by \citet[][dashed dotted lines,]{Arribas2014} is shown. Median of $\log_{10}(\mstar)$ are artificially displayed by $+0.05$ dex for the expelled mass rate and $-0.05$ dex for the unbound outflows, and the median of $\Vc$ is artificially moved +10 km/s for the expelled mass rate (orange circles) and -10 km/s for the unbound outflows (green circles), for visualisation purposes only.}
         \label{fig:eta_redshift}
\end{figure*}

We also added the median simulated $\eta$ in each figure. 
It is important to note that the comparison can only be done in a general way due to the low statistical number of data in both observations and simulations.  
In Fig.~\ref{fig:eta_redshift} we show the mass-loading factor as a function of $\Vc$ (upper panels) and  $\mstar$ (bottom panels) at $z \sim 1$ (left panels) and $z\sim 2$ (right panels). 
Observations from different authors have been included for comparison, as well as proposed fitting relations.

  Our estimations yield $\eta_{\rm out} \sim 0.6-4$ for the unbound outflows and  $\eta_{\rm ex} \sim 0.1-2$ for the expelled mass as displayed in Fig.~\ref{fig:eta_redshift} (see also Table~\ref{table:mass-loadingfactor}). For $z \sim 1$, sub-MW galaxies have a median of $\log_{10}(\eta_{\rm ex}) =  -0.10^{0.02}_{-0.24}$ dex for the expelled mass rate and of  $\log_{10}(\eta_{\rm out}) = 0.38^{0.48}_{0.10}$ dex for the unbound outflows, whereas for higher-mass galaxies the median value for the mass-loading factors for the expelled mass rate is $\log_{10}(\eta_{\rm ex}) =  0.16^{0.21}_{-0.10}$ dex and of $\log_{10}(\eta_{\rm out}) =  0.16^{0.17}_{0.12}$ dex for the unbound outflows.

  For the expelled mass the trends are very weak, and we find no significant correlation. The $\eta_{\rm out}$ obtained for the analysed \cielo~galaxies are  consistent with available observations of \citet{Schroetter2024} at  $z \sim 1$ with $\eta_{\rm out}^a \propto \Vc^{-0.48}$ for the unbound outflows. This correlation seems to be stronger at  $z\sim 2$ (upper-right panel), which yields $\eta_{\rm out}^a \propto \Vc^{-0.77}$ and of $\eta_{\rm ex} \propto \Vc^{-0.38}$ for the expelled mass (see Table~\ref{table:fitting_circular velocity} for the Spearman correlation factors). 
 Our fitted relations follow the observational data, suggesting that sub-MW galaxies are more affected by SN outflows than high-mass galaxies as expected. 
 
 Finally, in Fig.~\ref{fig:eta_redshift} (bottom panel), we study the dependence on $\mstar$. At $z\sim1$ we obtain a weak dependence on stellar mass at redshift  (bottom left panel) of $\eta_{\rm out}^a \propto \mstar^{-0.21}$, but statistically significant as can be seen from Table~\ref{table:fitting_circular velocity}. 
 The trends are consistent with observed galaxies \citep{Perrotta2023,Schroetter2024,Weldon2024}. Similarly to the trend with $\Vc$, we find that the dependence on $\mstar$ is stronger at $z\sim2$, which yields $\eta_{\rm out}^a \propto \mstar^{-0.30}$  and of $\eta_{\rm ex} \propto \mstar^{-0.37}$ (see also Table~\ref{table:fitting_circular velocity}).
 Our estimates agree with observations at the high-mass end  \citep{concas2022,Llerena2023,Weldon2024}. However, there are very limited observations for dwarf galaxies at this redshift. For $z \sim2$, sub-MW galaxies have a median of  $\log_{10}(\eta_{\rm ex}) =  -0.42^{-0.15}_{-0.78}$ dex for the expelled mass rate and of  $\log_{10}(\eta_{\rm out}) = -0.22^{0.16}_{-0.51}$ dex for the unbound outflows, whereas for higher-mass galaxies the median value for the mass-loading factors for the expelled mass rate is $\log_{10}(\eta_{\rm ex}) =  -0.86^{-0.68}_{-0.93}$ dex and of $\log_{10}(\eta_{\rm ex}) =  -1.00^{-0.95}_{-1.07}$ dex for the unbound outflows. For these figures, the median of $\log_{10}(\mstar)$ has been artificially displayed by $+0.05$ dex for the expelled mass rate and $-0.05$ dex for the unbound outflows, and the median of $\Vc$ is artificially moved +10 km/s for the expelled mass rate (orange circles) and -10 km/s for the unbound outflows (green circles), for visualisation purposes only.

 In summary, we observe that the mass-loading factor shows a clearer anti-correlation with $\Vc$ and $\mstar$ for higher redshift. This dependence is consistent with the fact that the SFR increases with redshift and galaxies became smaller and hence, more susceptible to being affected by SN feedback. As shown in Fig.~\ref{fig:SFH_all}, the frequency of mergers increases with redshift and their occurrences are also associated to the occurrences of starbursts, indicating a connection between them as suggested by observational and simulated works \citep[e.g.][]{tissera2001, kohandel2025}. As mentioned, the progenitors are also lower mass systems which makes them easier to trigger galactic outflows. We speculate these are the reasons behind the stronger correlations. As we move to lower redshift, the star formation diminishes and there are no clear correlations with mass or velocity at least for the \cielo~ galaxies.

\section{Conclusions}\label{sec:discussion and conclusions}
 We studied 15 central galaxies of the high-resolution CIELO-P7 cosmological simulations \citep{Tissera2025}, focussing on the sub-MW galaxies ($\mstar<10^{10}\Msun$). We followed the progenitor galaxies between redshift $z = [0,7]$ and measured unbound outflows in two shells of the CGM [1.5 $\ropt$, 0.5 $\rvir$] and $[0.5\rvir, \rvir)$, and expelled mass rates ($r > \rvir$).   We applied these two dynamically motivated definitions of outflows. \\
 We summarise our main results as follows:
   \begin{enumerate}
      \item We find that minor and major mergers and interactions, as well as SN feedback triggered by these events, can modulate the SFR histories of sub-MW galaxies. A tidally induced or merger-induced burst of SF triggers mass-loaded outflows, which decreases the star formation activity, as expected, due to the energy release of SN into the ISM, and transports enriched material out of a galaxy. However, we note that the impact of mergers on star formation and the galactic outflow is complicated and might depend on other parameters such as the gas-richness and the orbital characteristics of the encounters. A detailed analysis of these aspects is beyond the scope of this paper. 
     
      \item The outflow metallicity becomes significant relative to the ISM metallicity from $z \sim 2$, as many SF episodes have time to enrich a galaxy. For $z > 2$, we estimate a median $\rm Z_{out}/Z_{ISM} \sim 1.5$ for the sub-MW galaxies and a median $\rm Z_{out}/Z_{ISM} \leq 0.5$ for higher-mass galaxies,  in agreement with previous studies \citep[e.g.][]{Muratov2017,christensen2018}. At higher redshift ($z > 2$), the progenitors of the more massive systems have an average $Z_{\rm out}/Z_{\rm ISM} = 0.56$, while the sub-Milky-Way systems show a higher average ratio of $Z_{\rm out}/Z_{\rm ISM} = 0.67$. Sub-MW galaxies are more efficient in removing metals from their ISM than more massive systems across all redshift.

      \item The MZR of our sample is in agreement with observations. The effective yields show a large diversity in agreement with observations, particularly for sub-MW systems. There is a global agreement with predictions from analytical models that allow mass loss. Galaxies also have accreted gas that tends to have a lower metallicity than the star-forming ISM, particularly for sub-MW systems, although there is a large variation. The relative impact of inflows and outflows remains to be analysed. 
      
      \item Sub-MW galaxies have a higher fraction of their current total oxygen mass in the gas phase (i.e. ISM or CGM) and have higher fractions of expelled mass at $z = 0$ than higher-mass galaxies. This implies that sub-MW galaxies are more affected by outflows. We find that in galaxies with $\rm M_{*} < 10^{8.7}~\rm M_{\odot}$, the CGM contains between 10 and 40 percent of the oxygen mass within $\rvir$, whereas in more massive galaxies, this fraction drops below 10 percent. For sub-MW-mass galaxies, an equivalent to approximately $10–60$ percent of the present-day oxygen mass has been expelled into the IGM, while in more massive galaxies, this fraction is less than 10 percent. In addition, most of the oxygen in the CGM is in the form of low ion gas for $M_{*} < 10^{8.7}~\rm M_{\odot}$. This component seems to be a metal reservoir in the CGM of these galaxies. 
      
      \item We obtain a mass-loading factor between $\eta_{\rm out} \sim$ 0.7-6 for the unbound outflows and $\eta_{\rm ex} \sim 0.6-5.7$ for the expelled mass rate for $z \leq 0.5$. This is consistent with observations of local dwarf galaxies \citep{mcquinn2019} and field galaxies \citep{forster2019}. We find no clear signals of correlation between the mass-loading factor and the circular velocity in this redshift range. 

      \item At $z\sim 1$  we obtain mass-loading factors within the range $\eta_{\rm out} \sim$ 0.6-4 for the unbound outflows and $\eta_{\rm ex} \sim 0-2$ for the expelled mass rate, and for $z\sim 2$ between $\eta_{\rm out,ex} \sim 0-3$ for the unbound outflows and for the expelled mass rate. The negative correlation is stronger at higher redshift ($z\sim 2$) with a linear fit $\eta_{\rm out} \sim \Vc^{-0.71}$. This suggests that the progenitors of sub-MW galaxies are more affected by SN outflows than those of higher-mass galaxies. The negative correlation is stronger at $z\sim 2$ as progenitors of the selected galaxies are more actively forming stars, have more frequent minor and major mergers, and lower potential wells.   
   \end{enumerate}

The \cielo~galaxies are able to reproduce global observed trends of sub-MW galaxies and provide a reference to compare with observations and with simulations using different subgrid physics. Sub-MW galaxies provide important constraints to test subgrid modelling. We predict that a significant fraction of metals in low-mass galaxies are stored in the cool components of their CGM. Due to the high number abundance of dwarf galaxies, they could represent a significant reservoir of chemical elements.

\begin{acknowledgements}
        We thank the anonymous referee for their rigorous and insightful comments which helped improve the clarity of this manuscript. We gratefully thank the EvolGal4D team for the insightful discussions throughout this work. We also thank Brian Tapia-Contreras for developing the code used to produce Fig.1. VPM acknowledges funding by ANID (Beca Magíster Nacional, Folio 22241063). PBT acknowledges partial funding by Fondecyt-ANID 1240465/2024. JGJ acknowledges funding by ANID (Beca Doctorado Nacional, Folio 21210846). ES thanks partial financial support by Fondecyt-ANID Postdoctoral 2024 Project N°3240644. RDT thanks the Ministerio de Ciencia e Innovación (Spain)  for financial support under Project grant PID2021-PID2024-156100NB-C21  financed by MICIU/AEI /10.13039/501100011033 / FEDER, EU. We acknowledge ANID Basal Project FB210003 and  N\'ucleo Milenio ERIS. This work has received financial support from the European Union's HORIZON-MSCA-2021-SE-01 Research and Innovation programme under the Marie Sklodowska-Curie grant agreement number 101086388 - Project acronym: LACEGAL. This project used the Ladgerda Cluster (Fondecyt 1200703/2020 hosted at the Institute for Astrophysics, Chile), the NLHPC (Centro de Modelamiento Matem\'atico, Chile) and the Barcelona Supercomputer Center (Spain). This work made use of Astropy \citep{astropy2013} and Py-SPHViewer \citep{benitez2015}, we acknowledge the developers.
\end{acknowledgements}

\bibliographystyle{aa} 
\bibliography{biblio}

\begin{appendix} 

\section{The Tully-Fisher relation} \label{appendix:TFR}
 
The \cielo~galaxies follow the Tully-Fisher relation \citep[TFR - ][]{tullyfisher1977}  as shown by \citet{Tissera2025}.  In this paper, we estimated it again, but for dwarf galaxies. The main goal is to estimate the potential well of the galaxies by using the rotational velocity as a proxy. 
 
 For this purpose, we obtained the rotational curves of the gas component for each galaxy and their progenitors by estimating the tangential velocity of each gas particle. The maximum rotational velocity, V$_{\rm max}$, was measured as the mean tangential velocity within 0.5 kpc radial interval around the maximum value achieved. Additionally, we define the circular velocity, $\Vc$, as the rotational velocity measured at  $\ropt$. Both values are given in Table~\ref{table:physicalparameters}. When there is a well-defined gaseous disc in equilibrium within its potential well, we expect $\Vc = \sqrt{G \mstar /r}$. This is the case for the simulated galaxies with a disc. However, there are a few galaxies in which the gas components are not settled onto well-behaved discs, hence, the gas is more turbulent and dominated by dispersion. For these galaxies, we take $\Vc$ as the maximum value of the curve, $\Vmax$.

 \begin{figure}[h]
   \centering
   \includegraphics[width=\hsize]{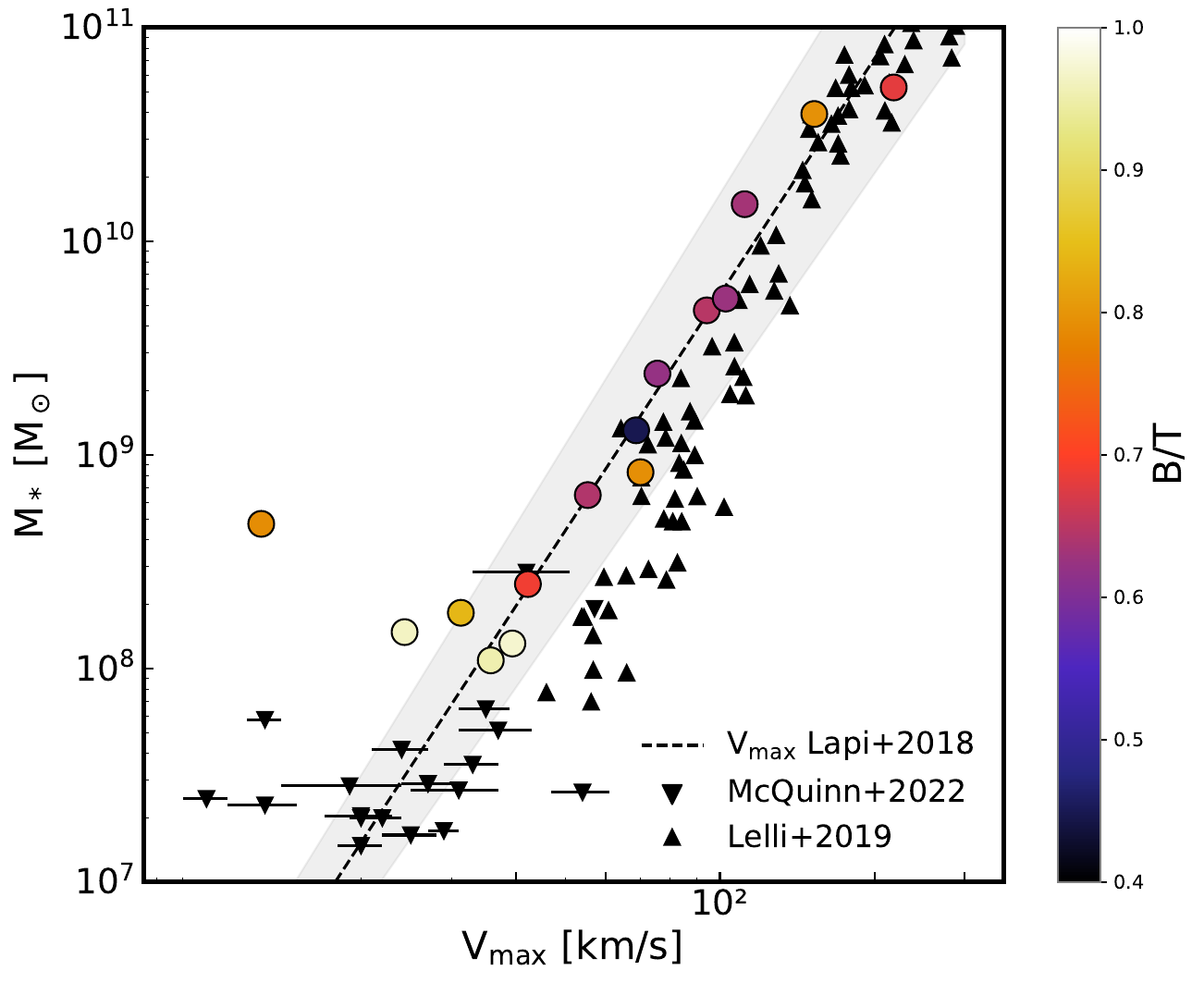}
      \caption{The Tully-Fisher relation obtained by using the maximum rotational velocity, $\Vmax$, for the \cielo~ galaxies, coloured by B/T. Observations by \citet[][downwards-pointing black triangles]{mcquinn2022} and \citet[][black triangles]{lelli2019} are displayed from comparison. The best-fitting regression for the data of \citet{lapi2018A} is also shown (black, dashed line). The shaded region is defined by $1\sigma$.}
         \label{fig:TFR}
\end{figure}

 Figure~\ref{fig:TFR} displays the TFR for the simulated galaxies, coloured according to the B/T ratio.
 For comparison, the best-fitting linear regression (dashed line) M$_{\rm bar}$ = 3.67 $\times$ $\Vmax$ + 2.41 determined by \citet{lapi2018A} of local spiral galaxies was included. 
 Additionally, we include observations of dwarf irregular galaxies by \citet{lelli2019}, where stellar masses were estimated by this author are based on 3.6 $\mu$m fluxes and velocities were measured at the peak of their rotational curves. Observational measurements of low-mass galaxies by \citet{mcquinn2022}, which used rotational velocities estimated by using PV slices (position-velocity diagram) of HI zones are also included.\\
 Our \cielo~galaxies follow the observed trend, except for very low-mass systems, which are mostly dominated by dispersion (0200, 2736, and 9110) and, hence, their $\Vmax$ are smaller than expected, since we only estimated the TFR to have a characterisation of the simulated galaxies, calculating a correction to the TFR such as $S_{0.5}$ is beyond the scope of this paper \citep{derossi2012}.

\section{The MZR for inflows} \label{appendix:MZR}

We present the MZR for the ISM (purple circles) and for the inflow around $z=0$ (cyan circles) for our \cielo~-P7 galaxies. Similar to Fig.~\ref{fig:mzr}, our simulated galaxies follow the MZR. We note that the inflows of these galaxies at $z=0$ is mainly metal-poor gas, however, higher-mass galaxies tend to have slightly more enriched gas than sub-MW galaxies. As mentioned in Sect.~\ref{sec:outflows}, every galaxy has a unique SFH; therefore, sub-MW galaxies have a diversity in the inflow metallicity.

\begin{figure}[h]
   \centering
   \includegraphics[width=\hsize]{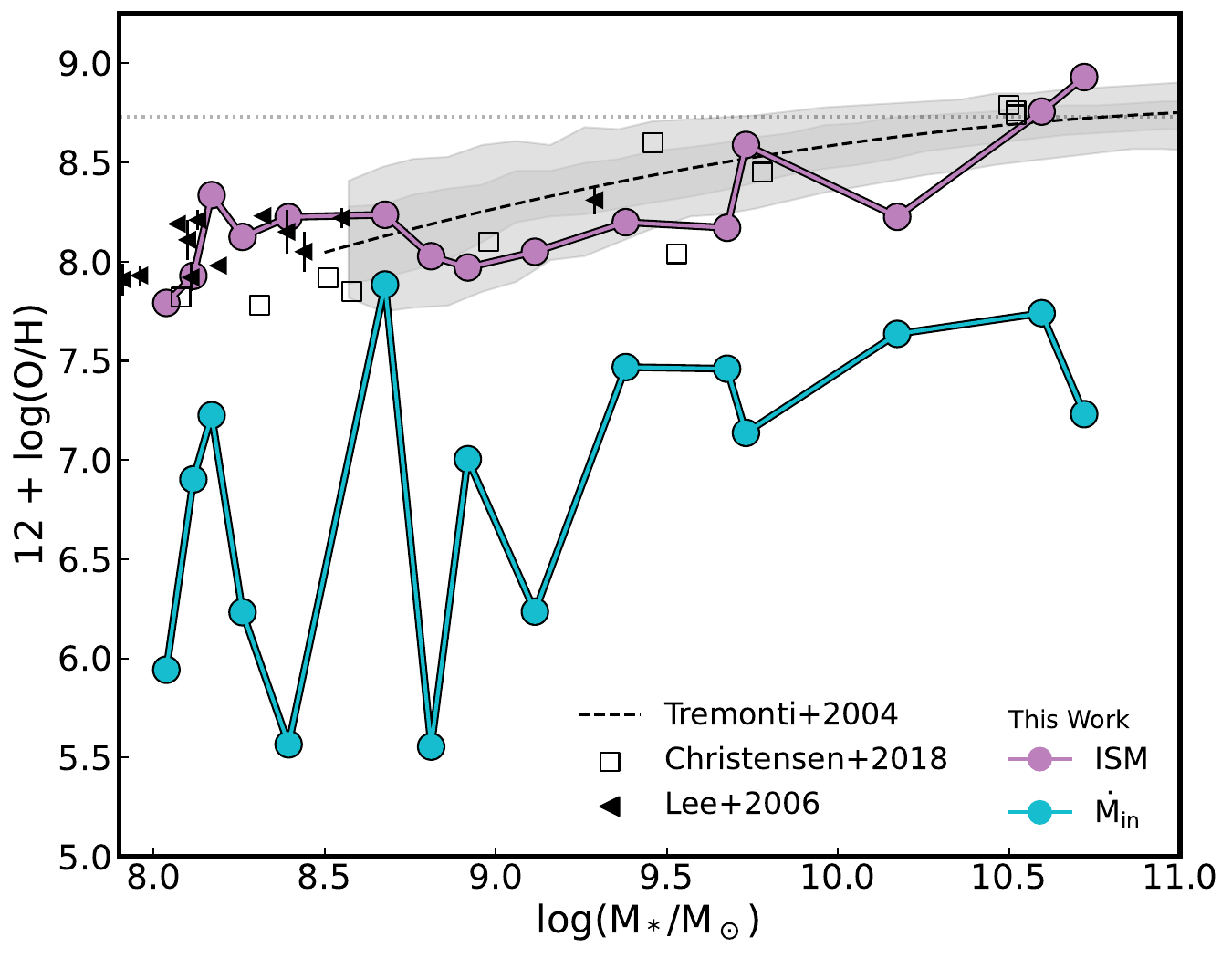}
      \caption{Mass-metallicity relation for the ISM (purple circles) and for the inflow rate around $z=0$ (cyan circles). Symbols are described in Fig.~\ref{fig:mzr}.}
         \label{fig:MZR_in}
\end{figure}

\onecolumn
\section{Outflows for~\cielo-P7 galaxies} \label{appendix:outflows}

\begin{figure}[h]
      \includegraphics[width=\hsize]{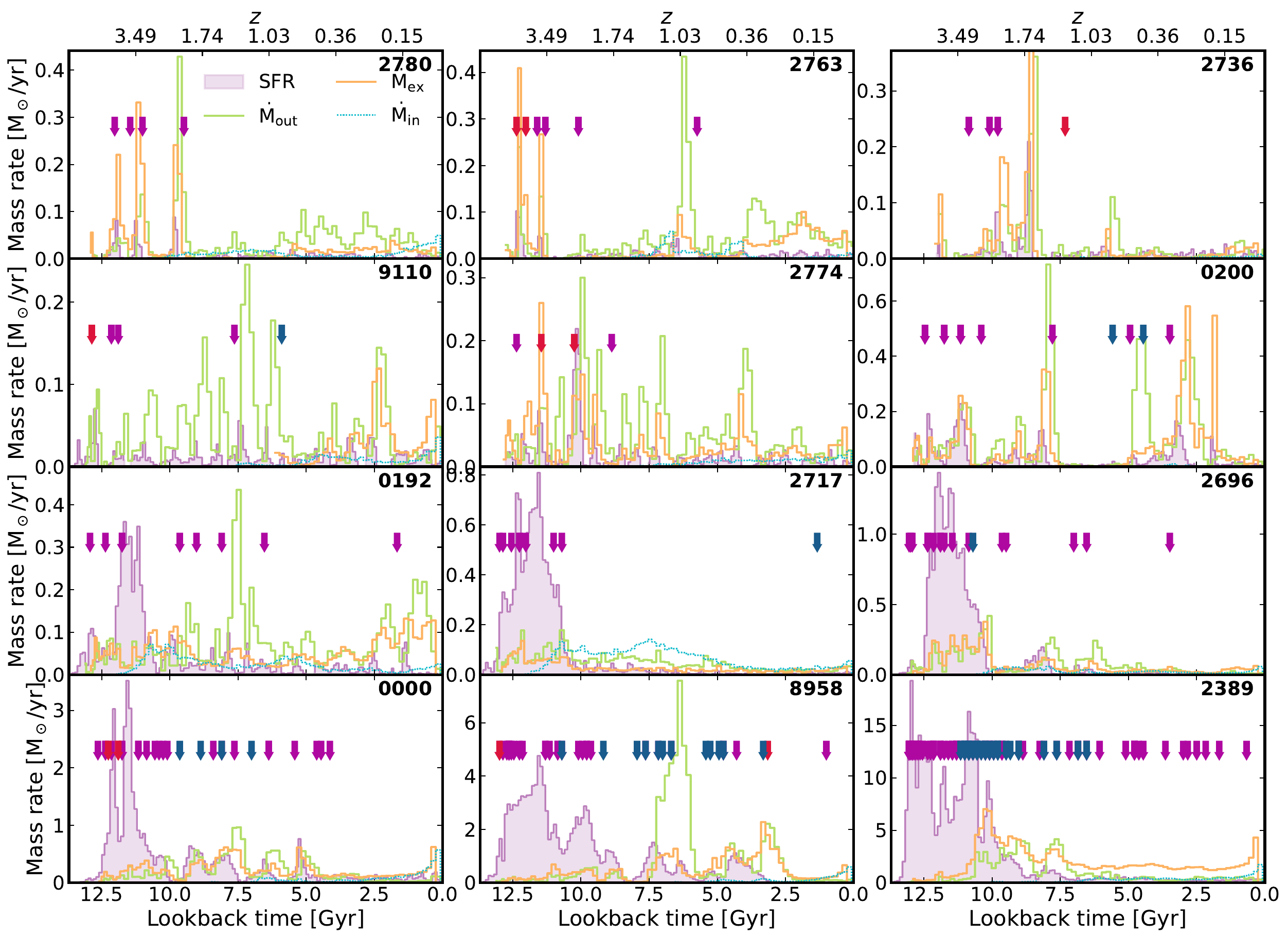}
      \caption{Evolution of SFR (purple shades), $\mout$, the rate of unbound outflow for the inner shell [1.5$\ropt$, 0.5$\rvir$) (solid, green lines), $\mex$, the expelled mass rates (solid, orange lines), and $\dot{M_{\rm in}}$, the inflow mass rate (dotted, cyan line) as a function of lookback time. The infall time of satellites (blue arrows) entering the virial radius, and the time of minor and major mergers (pink and red arrows, respectively) when corresponds.}
         \label{fig:SFH_all}
\end{figure}

Figure~\ref{fig:SFH_all} shows the unbound outflows, expelled mass rates, SFRs, and inflow rates as described in Sect.~\ref{sec:outflows} for the remaining~\cielo-P7 galaxies.

\newpage
\section{Summary of main parameters.} \label{appendix:parameters}

Here we present a summary of the physical properties and main parameters obtained for our sample of \cielo-P7 galaxies summarised in Tables \ref{table:galaxy properties}, \ref{table:oxygen fraction}, \ref{table:fitting_circular velocity} and  \ref{table:mass-loadingfactor}.
In the case of \ref{table:mass-loadingfactor} we also provide the Spearman coefficients so the reader can use the fitted relations with caution depending on the level of statistical significance.\\

\begin{table*}[h!]     
\caption{Physical properties of the selected \cielo~galaxies at $z = 0$.}
\label{table:galaxy properties}      
\centering                      
\begin{tabular}{c c c c c}     
\hline\hline             
Galaxy ID & $\Sigma_{\rm gas}$ &$\Sigma_{*}$ & log(T$_{\rm vir}$/K) & $\mu$ \\   
 & $\Msun$/pc$^2$ &$\Msun$/pc$^2$& & \\
\hline                     
   2780 & 0.73 & 0.19 & 3.95 & 0.87\\
   2763 & 0.54 & 0.24 & 4.09&  0.80\\
   2736 & 0.73 & 0.98 & 3.91&  0.58 \\
   9110 & 0.51 & 0.57 & 3.97&  0.68 \\
   2774 & 1.23 & 1.64 & 4.01&  0.58\\
   0200 & 0.31 & 2.60 & 4.16&  0.26 \\
   0192 & 1.46 & 1.10 & 4.21&  0.63 \\
   0181 & 1.78 & 0.53 & 4.25&  0.77 \\
   2717 & 1.51 & 1.17 & 4.31& 0.63 \\  
   2696 & 1.10 & 3.93 & 4.48& 0.34 \\
   2627 & 1.87 & 3.02 & 4.74& 0.52 \\
   0000 & 1.91 & 9.75 & 4.69& 0.34 \\
   8958 & 1.51 & 18.96 &  4.920& 0.20 \\
   7805 & 0.76 & 41.63 & 5.32& 0.28 \\
   2389 & 2.35 & 62.82 & 5.31& 0.23\\
\hline    
\end{tabular}
\tablefoot{Columns from left to right summarise the gas surface density, stellar surface density, virial temperature, and gas fraction of the analysed galaxies.}
\end{table*}

\begin{table*}[h!]  
\caption{Oxygen fraction per component for CIELO galaxies at $z = 0$.}
\label{table:oxygen fraction}   
\centering                  
\begin{tabular}{c c c c c c c c c c c}       
\hline\hline
Galaxy ID & $\mstar$/M$_{\rm O,gx}$ & M$_{\rm g}$/M$_{\rm O,gx}$ & M$_{\rm ex}$/M$_{\rm O,gx}$ & M$_{\rm in}$/M$_{\rm O,gx}$ & M$_{\rm H}$/M$_{\rm O,CGM}$ & M$_{\rm W}$/M$_{\rm O,CGM}$ & M$_{\rm L}$/M$_{\rm O,CGM}$ & M$_{\rm C}$/M$_{\rm O,CGM}$\\ 
\hline  
   2780 & 0.15 & 0.85 & 0.17 & 0.06 & 0.00 & 0.00 & 0.68 & 0.31 \\
   2763 & 0.20 & 0.80 & 0.30 & 0.08 & 0.00 & 0.00 & 0.67 & 0.33 \\
   2736 & 0.31 & 0.69 & 0.20 & 0.02 & 0.00 & 0.00 & 0.71 & 0.29 \\
   9110 & 0.31 & 0.69 & 0.38 & 0.03 & 0.00 & 0.00 & 0.67 & 0.33 \\
   2774 & 0.45 & 0.55 & 0.36 & 0.03 & 0.00 & 0.01 & 0.80 & 0.19 \\
   0200 & 0.77 & 0.23 & 0.63 & 0.01 & 0.00 & 0.01 & 0.69 & 0.30 \\
   0192 & 0.79 & 0.31 & 0.15 & 0.04 & 0.00 & 0.01 & 0.76 & 0.23 \\
   0181 & 0.34 & 0.66 & 0.10 & 0.06 & 0.00 & 0.09 & 0.82 & 0.09 \\
   2717 & 0.73 & 0.27 & 0.04 & 0.01 & 0.00 & 0.01 & 0.82 & 0.17 \\      
   2696 & 0.89 & 0.11 & 0.05 & 0.01 & 0.00 & 0.02 & 0.71 & 0.26 \\
   2627 & 0.73 & 0.27 & 0.13 & 0.03 & 0.10 & 0.05 & 0.70 & 0.17 \\
   0000 & 0.83 & 0.17 & 0.05 & 0.01 & 0.12 & 0.05 & 0.71 & 0.12 \\
   8958 & 0.96 & 0.04 & 0.06 & 0.00 & 0.06 & 0.05 & 0.73 & 0.17 \\
   7805 & 0.91 & 0.09 & 0.07 & 0.02 & 0.44 & 0.04 & 0.46 & 0.06 \\
   2389 & 0.93 & 0.07 & 0.04 & 0.01 & 0.67 & 0.02 & 0.29 & 0.02 \\
\hline       
\end{tabular}
\tablefoot{Columns from left to right contain the oxygen fraction in the stellar component and the gas component of the galaxy's mass until $z = 0$, expelled and inflow oxygen fraction across $z = 0-7$ to the galaxy's total oxygen mass at $z = 0$, oxygen fraction of the CGM phase in forms of hot phase, warm ions, low ions, and cold phase.}
\end{table*}

\begin{table*}[h!]
\caption{Mass-loading factor measured in a redshift bin of $z \sim 0 $, $z\sim1$ and $z\sim2$.}
\label{table:mass-loadingfactor}   
\centering                  
\begin{tabular}{c c c c c c c c c c}      
\hline\hline
 & & $z \sim 0 $ & & & $z \sim 1$& & &  $z \sim 2$ & \\ \cline{2-10}
Galaxy ID & $\eta_{\rm out}^a$ & $\eta_{\rm out}^b$ & $\eta_{\rm ex}$&   $\eta_{\rm out}^a$ & $\eta_{\rm out}^b$ & $\eta_{\rm ex}$ & $\eta_{\rm out}^a$ & $\eta_{\rm out}^b$ & $\eta_{\rm ex}$\\
\hline  
   2780 & 6.04 & 6.02 & 2.30 & 3.33 & 3.25 & 0.50 & 2.61 & 1.91 & 3.04\\
   2763 & 4.32 & 4.32 & 2.78 & 3.86 & 3.65 & 0.99 & 1.83 & 1.71 & 0.16\\
   2736 & 0.67 & 0.67 & 0.57 & 1.83 & 1.84 & 1.98 & 0.62 & 0.84 & 1.24\\
   9110 & 1.65 & 1.64 & 1.53 & 4.00 & 3.41 & 0.11 & 2.90 & 2.66 & 0.00\\
   2774 & 2.55 & 2.52 & 1.67 & 2.96 & 2.59 & 1.23 & 1.27 & 1.25 & 0.65\\
   0200 & 2.92 & 2.91 & 2.43 & 2.62 & 2.58 & 1.42 & 1.33 & 1.24 & 0.91\\   0192 & 3.76 & 3.81 & 3.09 & 2.72 & 2.65 & 0.99 & 0.29 & 0.34 & 0.50\\
   0181 & 2.01 & 2.02 & 0.94 & 1.22 & 1.53 & 0.49 & 0.59 & 0.93 & 0.47\\
   2717 & 3.09 & 2.88 & 2.10 & 2.29 & 2.23 & 0.82 & 0.36 & 0.34 & 0.17\\ 
   2696 & 2.58 & 2.29 & 3.10 & 1.27 & 1.18 & 0.73 & 0.32 & 0.29 & 0.30\\
   2627 & 1.73 & 1.72 & 2.44 & 0.66 & 0.64 & 0.60 & 0.25 & 0.25 & 0.16\\
   0000 & 0.66 & 0.68 & 0.99 & 0.88 & 0.78 & 0.75 & 0.27 & 0.24 & 0.29\\
   8958 & 1.70 & 1.77 & 1.98 & 1.55 & 1.61 & 0.44 & 0.10 & 0.08 & 0.14\\
   7805 & 3.73 & 3.83 & 5.72 & 1.19 & 1.11 & 1.45 & 0.07 & 0.05 & 0.10\\
   2389 & 0.81 & 1.06 & 3.34 & 1.44 & 1.28 & 1.83 & 0.13 & 0.09 & 0.29\\
\hline                                
\end{tabular}
\tablefoot{$\eta_{\rm out}$ are mass-loading factors for the unbound outflows measured in a radial bin $^a$: [1.5$\ropt$,0.5$\rvir$) and $^b$: [0.5$\rvir$,$\rvir$).  $\eta_{\rm ex}$ are mass-loading factors for the expelled mass rate.}
\end{table*}

\begin{table*}[h!]  
\caption{Best fitting parameters for the mass-loading factors.}             
\label{table:fitting_circular velocity}      
\centering          
\begin{tabular}{c c c c c c c c c c c c c c c }     
\hline\hline

&&&&$\Vc$&&&&&$\mstar$&\\
 & & $\eta_{100}$ & $\beta$ & $\sigma_{\eta_{100}}$ & $\sigma_{\beta}$ & r   & $\eta_*$ & $\beta$ & $\sigma_{\eta_{*}}$ & $\sigma_{\beta}$ & r\\ 
\hline

       &$\eta_{\rm out}^a$ &2.26 & 0.30 & 0.45 & 0.27 & -0.24 (0.390) & 3.22 & 0.11 & 0.63 & 0.08 & -0.28 (0.315)\\
   $z \leq 0.5$ &$\eta_{\rm out}^b$ & 2.28 &  0.28 & 0.46 & 0.27 & -0.20 (0.475) &3.17 &  0.10 & 0.62 & 0.08 & -0.24 (0.398) \\
   &$\eta_{\rm ex}$    & 2.62 & -0.46 & 0.30 & 0.19 & 0.47 (0.079) & 1.57 & -0.15 & 0.36 & 0.05 & 0.44 (0.104)\\
\hline                       
   &$\eta_{\rm out}^a$ & 1.61 & 0.48 & 0.27 & 0.16 & -0.66 (0.007)& 2.92 & 0.21 & 0.23 & 0.05 & -0.79 (0.001)\\  
   $z \sim 1$ &$\eta_{\rm out}^b$ & 1.55 &  0.47 & 0.24 & 0.14 & -0.68 (0.005) & 2.76 &  0.20 & 0.20 & 0.04 & -0.83 (0.000)\\
   &$\eta_{\rm ex}$    & 0.99 & -0.09 & 0.16 & 0.20 & -0.09 (0.761)& 0.83 & -0.07 & 0.19 & 0.07 & 0.11 (0.694)\\
\hline 
   &$\eta_{\rm out}^a$ & 0.50 & 0.77 & 0.23 & 0.33 & -0.84 (0.001)&1.09 & 0.30 & 0.21 & 0.11 & -0.89 (0.001)\\  
   $z \sim2$ &$\eta_{\rm out}^b$ & 0.47 &  0.76 & 0.19 & 0.28 & -0.90 (0.001) & 1.03 &  0.29 & 0.16 & 0.09 & -0.89 (0.001)\\
   &$\eta_{\rm ex}$    & 0.64 & 0.38 & 0.25 & 0.49 & -0.35 (0.206) & 0.64 & 0.37 & 0.23 & 0.21 & -0.47 (0.079)\\
\hline
\end{tabular}
\tablefoot{$\eta$ = $\eta_{100}$($\Vc$/${100 \rm km s^{-1}}$)$^{- \beta}$ and $\eta$ = $\eta_*$($\mstar$/$10^8\Msun)^{- \beta}$ for the expelled mass ($\eta_{\rm ex}$), and for the unbound outflows measured in a radial bin $^a$: [1.5R$_{\rm opt}$,0.5R$_{\rvir}$) and $^b$: [0.5$\rvir$,$\rvir$). Columns from left to right contain $\eta_{100}$($\eta_*$), $\beta$, standard deviations $\sigma_{\eta_{100}}$($\sigma_{\eta_*}$), $\sigma_\beta$, and the Spearman coefficients r and p (within parenthesis).}
\end{table*}

\end{appendix}

\label{LastPage}
\end{document}